\documentclass{acmtrans2m}
\usepackage{amsmath,graphicx,amsxtra}
\usepackage{psfrag}
\usepackage{amsfonts, amssymb}
\usepackage{eurosym}

\def\0{{\bf 0}}
\def\1{{\bf 1}}

\newtheorem{theorem}{Theorem}[section]

\newtheorem{corollary}{Corollary}[section]
\newtheorem{definition}{Definition}[section]

\title{Unicast and Multicast QoS Routing with Soft Constraint Logic Programming}

\author{STEFANO BISTARELLI \\ Universit\`a
           Chieti-Pescara, Istituto di Informatica e Telematica  \\ UGO MONTANARI\\ Universit\`a di Pisa \\ FRANCESCA ROSSI\\Universit\`a  di Padova \and FRANCESCO SANTINI \\ IMT  - Institute for Advanced Studies, Istituto di Informatica e Telematica}

\begin{abstract}
We present a  formal model to represent and solve the
unicast/multicast routing problem in networks with \emph{Quality
of Service} (QoS) requirements. To attain this, first we translate
the network adapting it to a weighted graph (unicast) or
\emph{and-or} graph (multicast), where the weight on a connector
corresponds to the multidimensional cost of sending a packet on
the related network link: each component of the weights vector
represents a different QoS metric value (e.g. bandwidth, cost,
delay, packet loss). The second step consists in writing this
graph as a program in \emph{Soft Constraint Logic Programming}
(\emph{SCLP}): the engine of this framework is then able to find
the best paths/trees by optimizing their costs and solving the
constraints imposed on them (e.g. $delay \leq 40msec$), thus
finding a solution to QoS routing problems. Moreover,
\emph{c-semiring} structures are a convenient tool to model QoS
metrics. At last, we provide an implementation of the framework
over scale-free networks and we suggest how the performance can be
improved.
\end{abstract}






\begin{document}

\begin{bottomstuff}
Author's address: Stefano Bistarelli,  Dipartimento di Scienze,
Universit\`{a} di Chieti-Pescara, Viale Pindaro 42, Pescara,
65127, Italy. {\tt bista@sci.unich.it.} - Institute for
Informatics and Telematics, Via G. Moruzzi 1, 56100 Pisa, Italy.
{\tt stefano.bistarelli@iit.cnr.it.}
\newline Ugo Montanari, Dipartimento di Informatica, Universit\`{a} di
Pisa, Largo Bruno Pontecorvo 3, Pisa, 56127, Italy. {\tt
ugo@di.unipi.it}
\newline Francesca Rossi, Dipartimento di Matematica Pura e Applicata, Universit\`{a} di Padova,  Via Trieste 63
35121 Padova, Italy. {\tt frossi@math.unipd.it}
\newline Francesco Santini, IMT - Institute for Advanced Studies, Piazza San Ponziano 6, 55100
Lucca, Italy. {\tt f.santini@imtlucca.it.} - Institute for
Informatics and Telematics, Via G. Moruzzi 1, 56100 Pisa, Italy.
{\tt francesco.santini@iit.cnr.it.}
\end{bottomstuff}

\maketitle

\section{Introduction}\label{sec:introduction}
Towards the second half of the nineties, Internet Engineering Task
Force (IETF) and the research community have proposed many service
models and mechanisms~\cite{bigPicture,qosnewsurvey} to meet the
demand for network \emph{Quality of Service} (QoS). The reason is
that traditional networks cannot recognize a priority associated
with data, because they handle network traffic with the \emph{best
effort} principles. According to this treatment, the network does
not provide any guarantees that data is delivered or that a user
is assisted with a guaranteed QoS level or a certain priority (due
to congestions). In best effort networks, all users obtain exactly
the same treatment. However nowadays, networked applications, such
as Enterprise Resource Planning (ERP), data mining, distance
learning, resource discovery, e-commerce, and distribution of
multimedia-content, stock quotes, and news, are bandwidth hungry,
need a certain ``timeliness'' (i.e. events occurring at a suitable
and opportune time) and are also mission critical.

For all these reasons, the routing problem has naturally been
extended to include and to guarantee QoS
requirements~\cite{CBR,bigPicture,qosnewsurvey}, and consequently
is usually abbreviated to \emph{QoS routing}. As defined
in~\cite{RFC2386}, QoS is ``a set of service requirements to be
met by the network while transporting a flow'', where a flow is
``a packet stream from source to a destination (unicast or
multicast) with an associated Quality of Service (QoS)''. To be
implemented and subsequently satisfied, service requirements have
to be expressed in some measurable QoS metrics, such as
bandwidth, number of hops, delay, jitter, cost and loss
probability of packets.

This paper combines and extends the two works presented
in~\cite{sclpjheu} and~\cite{qapl2007}. First, we detail the
modelling procedure to represent and solve plain Shortest Path
(SP)~\cite{Cormen} problems with \emph{Soft Constraint Logic
Programming} (see Sec.~\ref{sec:softCLP}). We consider several
versions of SP problems, from the classical one to the
\emph{multi-criteria} case (i.e. many costs to be optimized), from
\emph{partially-ordered} problems to those that are based on
modalities associated to the use of the arcs (i.e.
\emph{modality-based}), and we show how to model and solve them
via SCLP programs. The basic idea is that the paths represent
network routes, edge costs represent QoS metric values, and our
aim is to guarantee the requested QoS on the found unicast routes,
by satisfying the QoS constraints and optimizing the cost of the
route at the same time. The different criteria can be, for
example, maximizing the global bandwidth and minimizing the delay
that can be experienced on a end-to-end communication.

Then, extending the unicast solution, we suggest a formal model to
represent and solve the multicast routing problem in multicast
networks (i.e. networks supporting the multicast delivery schema)
that need QoS support. To attain this, we draw the network
adapting it to a weighted \emph{and-or} graph~\cite{mm78}, where
the weight on a connector corresponds to the cost of sending a
packet on the network link modelled by that connector. Then, we
translate the hypergraph in a SCLP program and we show how the
semantic of this program computes the best tree in the
corresponding \emph{and-or} graph. We apply this result to find,
from a given source node in the network, the multicast
distribution tree having the minimum cost and reaching all the
destination nodes of the multicast communication. The costs of the
connectors can be described as vectors (multidimensional costs),
each component representing a different QoS metric value. We show
also how modalities can be added  to multicast problems, and how
the computational complexity of this framework can be reduced.
Therefore, in this paper we present a complete formal model to
represent and solve the unicast/multicast QoS routing problem.

SCLP programs are logic programs where each ground atom can be
seen as an instantiated soft constraint~\cite{ijcai95,jacm97} and
it can be associated with an element taken from a set. Formally,
this set is a \emph{c-semiring}~\cite{bistabook} (or simply
semiring in the following), that is, a set plus two operations,
$+$ and $\times$, which basically say how to combine constraints
and how to compare them. The presence of these two operations
allows to replace the usual boolean algebra for logic programming
with a more general algebra where {\em logical and} and {\em
logical or} are replaced by the two semiring operations. In this
way, the underlying logic programming engine provides a natural
tool to specify and solve combinatorial problems, while the soft
constraint machinery provides greater expressivity and
flexibility.

The most important features of the adopted framework are: first,
is that SCLP is a declarative programming environment and, thus,
is relatively easy to specify a lot of different problems, ranging
from paths to trees. The model can be used to easily specify the
problem, which can be then translated and solved with a fast
solver; however, our goal is to improve the performance also for
our implementation. The second reason is that the semiring
structure is a very flexible and parametric tool where to
represent several and different cost models, with respect to QoS
metrics; obviously, the same SCLP programming environment and
operational semantic engine can be used with all these different
semirings. Finally, since QoS routing problem can be in general
NP-Complete, SCLP promises to be suitable tool, due to its ability
for solving combinatorial problems (as shown in~\cite{yan-cp98}).

\subsection{Related Works}\label{sec:relatedw}
Concerning the related works, in~\cite{programQoS}
and~\cite{shreq} the authors adopt a hypergraph model in joint
with semirings too, but the minimal path between two nodes (thus,
not over an entire tree) is computed via a graphical calculous
instead of SCLP. At the moment, all these frameworks are not
comparable from the computational performance point of view, since
they have  not yet been implemented. Even the work
in~\cite{mammeri} presents some general algebraic operators in
order to handle QoS in networks, but without any practical
results. We compare our work only with other theoretical frameworks, since our study
aims at representing general routing constraints in order to solve different problems: due to the
complexity of QoS routing, state-of-the-art practical solutions (presented in Sec.~\ref{sec:unicastbg}
and Sec.~\ref{sec:multicastbg}) deal only with a subset of metrics and constraints. On the other hand,
a more general framework can help to analyze the problem from a global point of view, not linked to specific algorithms.
With \emph{Declarative routing}~\cite{decrouting}, a routing protocol is implemented
by writing a simple query in a declarative query language (like
Datalog as in~\cite{decrouting}), which is then executed in a distributed fashion at
some or all of the nodes. It is based on the observation that recursive query languages
are a \emph{natural-fit} for expressing routing protocols. However, the authors of~\cite{decrouting} did not go deep in
modelling QoS features, and we think that c-semirings represent a very good method to include these metrics.

To go further, aside the elegant formalization due to the
SCLP framework, we build a bridge to a real implementation of the
model (Sec.~\ref{sec:implementation}) and several ideas to improve
the experienced performance. The final tool can be used to quickly prototype and test different routing paths.
Therefore, our paper vertically cover the problem: from theoretical to practical aspects, without reaching the performance of
existing routing algorithms implemented inside the routers, but thoroughly and expressively facing the problem.
As far as we know, other formal representations completely miss this practical implementation.

\subsection{Structure of the paper}\label{sec:paperstruct}

The remainder of this paper is organized as follows. In
Sec.~\ref{sec:softCLP} we describe the SCLP framework, while in
Sec.~\ref{sec:qosrouting} we complete the background by
introducing the multicast/unicast QoS routing: we show that the
problem of defining a route that has to be optimized and is
subject to constraints concerning QoS metrics, is, in general, a
NP-Complete problem. Then, we report some of the solutions, mostly
through heuristics, given in the real world.
Section~\ref{sec:unicastmodel} proposes how to model and solve the
unicast QoS routing with SCLP, considering also problems with
multidimensional costs (i.e. multi-criteria problems) and based on
modalities of use associated with the links of the network: for
example, if we need to find a route by using only wireless, and/or
wired and/or encrypted links (i.e. modality-based problems).
Section~\ref{sec:multicastmodel} outlines a similar framework,
based on hypergraph and SCLP, for the management of the multicast
QoS routing: we show how to translate a network in a corresponding
\emph{and-or} graph and then we compute the best distribution tree
by using SCLP. Even in this case we extend the model to include
problems with modalities. Section~\ref{sec:partiallyref} gives
some important considerations about semirings that improve the
model when the costs of the network links are multidimensional and
partially ordered: this is the common case, since an effective
measurement of QoS will necessarily involve a collection of
measures. We show also how we can limit the number of partially
ordered solutions with ad-hoc semirings, which apply a total order
on the tuples of cost values by following a set of weights defined
to satisfy the user. Section~\ref{sec:solving} presents a
practical implementation of the model by solving the problem over
\emph{scale-free}~\cite{barabasi} networks, which properly model
the topology of Internet. This implementation has been developed
to demonstrate that performance improvements are necessary. These
improvements can be achieved with the mechanisms explained in
Sec.~\ref{sec:complexity}, as \emph{tabling} and \emph{branch-and-bound} (as our implementation in \emph{ECLiPSe}~\cite{eclipse} shows).
At last, Sec.~\ref{sec:conclusions} ends the paper with the
final conclusions and ideas about future work.

\section{Soft Constraint Logic Programming}
\label{sec:softCLP}

The SCLP framework~\cite{bistabook,ijcai97,yan-cp98}, is based on
the notion of {\em c-semiring}  introduced
in~\cite{ijcai95,jacm97} (c-semiring and semiring terms will be
used as synonyms in this paper). A semiring $S$ is a tuple
$\langle A,+,\times, \0, \1 \rangle$ where $A$ is a set with two
special elements ($\0, \1 \in A$) and with two operations $+$ and
$\times$ that satisfy certain properties: $+$ is defined over
(possibly infinite) sets of elements of $A$ and thus is
commutative, associative, idempotent, it is closed and $\0$ is its
unit element and $\1$  is its absorbing element; $\times$ is
closed, associative, commutative, distributes over $+$, $\1$  is
its unit element, and $\0$ is its absorbing element (for the
exhaustive definition, please refer to~\cite{jacm97}).

The $+$ operation defines a partial order $\leq_S$ over $A$ such
that $a \leq_S b$ iff $a+b = b$; we say that $a \leq_S b$ if $b$
represents a value {\em better} \/than $a$. Other properties
related to the two operations are that $+$ and $\times$ are
monotone on $\leq_S$, $\0$ is its minimum and $\1$ its maximum,
$\langle A,\leq_S \rangle$ is a complete lattice and $+$ is its
lub. Finally, if $\times$ is idempotent, then $+$ distributes over
$\times$, $\langle A,\leq_S \rangle$ is a complete distributive
lattice and $\times$ its glb.

\emph{Semiring-based Constraint Satisfaction Problems}
(SCSPs)~\cite{bistabook} are constraint problems where each
variable instantiation is associated to an element of a c-semiring
$A$ (to be interpreted as a cost, level of preference, \ldots),
and constraints are combined via the $\times$ operation and
compared via the $\leq_S$ ordering. Varying the set $A$ and the
meaning of the $+$ and $\times$ operations, we can represent many
different kinds of problems, having features like fuzziness,
probability, and optimization. Notice also that the cartesian
product of two c-semirings is a c-semiring~\cite{jacm97}, and this
can be fruitfully used to describe multi-criteria constraint
satisfaction and optimization problems.

\emph{Constraint Logic Programming} (CLP)~\cite{clp-survey}
extends Logic Programming by replacing term equalities with
constraints and unification with constraint solving. The SCLP
framework extends the classical CLP formalism in order to be able
to handle also SCSP~\cite{ijcai95,jacm97} problems. In passing
from CLP to SCLP languages, we replace classical constraints with
the more general SCSP constraints where we are able to assign a
{\em level of preference} \/to each instantiated constraint (i.e.\
a ground atom). To do this, we also modify the notions of
interpretation,
model, model intersection, and others, 
since we have to take into account the semiring operations and not
the usual CLP operations.

The fact that we have to combine several refutation paths when we
have a partial order among the elements of the semiring (instead
of a total one), can be fruitfully used in the context of this
paper when we have an graph/hypergraph problems with incomparable
costs associated to the edges/connectors. In fact, in the case of
a partial order, the solution of the problem of finding the best
path/tree should consist of all those paths/trees whose cost is
not ``dominated'' by others.

\begin{table}[htb]
\caption{A simple example of an SCLP program.} \label{tab:sclp}
\hrule \vskip3pt
\begin{tabbing} {\tt p(a,b)} \= {\tt :- q(a).}
\kill
{\tt s(X) } \> {\tt :- p(X,Y).}\\
{\tt p(a,b)} \> {\tt :- q(a).}\\
{\tt p(a,c)} \> {\tt :- r(a).}\\
{\tt q(a)} \> {\tt  :- t(a).}\\
{\tt t(a)} \> {\tt  :- 2.}\\
{\tt r(a)} \> {\tt :- 3.}
\end{tabbing}
\vskip1pt \hrule
\end{table}

A simple example of an SCLP program over the semiring $\langle
\mathbb{N}, min,+,+\infty,0\rangle$, where $\mathbb{N}$ is the set
of non-negative integers and $D = \{a,b,c\}$, is represented in
Table~\ref{tab:sclp}. The choice of this semiring allows us to
represent constraint optimization problems where the semiring
elements are the costs for the instantiated atoms. To better
understand this Table, we briefly recall the SCLP syntax: a
program is a set of clauses and each clause is composed by a head
and a body. The head is just an atom, and the body is either a
collection of atoms, or a value of the semiring, or a special
symbol ($\Box$) to denote that it is empty. Clauses where the body
is empty or it is just a semiring element are called facts and
define predicates which represent constraints. When the body is
empty, we interpret it as having the best semiring element (that is, 1).  

The intuitive meaning of a semiring value like $3$ associated to
the atom $r(a)$ (in Table~\ref{tab:sclp}) is that $r(a)$ costs $3$
units. Thus the set $\mathbb{N}$ contains all possible costs, and
the choice of the two operations $min$ and $+$ implies that we
intend to minimize the sum of the costs. This gives us the
possibility to select the atom instantiation which gives the
minimum cost overall. Given a goal like $s(x)$ to this program,
the operational semantics collects both a substitution for $x$ (in
this case, $x=a$) and also a semiring value (in this case, $2$)
which represents the minimum
cost among the costs for all derivations for $s(x)$. 
To find one of these solutions, it starts from the goal and uses
the clauses as usual in logic programming, except that at each
step two items are accumulated and combined with the current
state: a substitution and a semiring value (both provided by the
used clause). The combination of these two items with what is
contained in the current goal is done via the usual combination of
substitutions (for the substitution part) and via the
multiplicative operation of the semiring (for the semiring value
part), which in this example is $+$. Thus, in the example of goal
$s(X)$, we get two possible solutions, both with substitution
$X=a$ but with two different semiring values: $2$ and $3$. Then,
the combination of such two solutions via the $min$ operation give
us the semiring value $2$.

To extend the representation we briefly introduce \emph{semiring
valuations}~\cite{nic}, which are constraint satisfaction problems
taking values in a commutative semiring, where the ordering is the
transitive relation $a \leq b$ iff $\exists.a + c = b$. The lack
of idempotency for the sum operator results in a weaker structure
than absorptive semirings, that has proved useful whenever
counting the number of solutions is of interest. Even if
throughout the paper we will use the semiring definition given
in~\cite{ijcai95,jacm97} (i.e. with an idempotent $+$), semiring
valuations can be used for those metrics that need to be
aggregated together from different solutions, e.g. the packet loss
probability between two nodes $p$ and $v$ is computed with the
probabilities of all the possible different paths connecting $p$
and $v$ in the graph. However, the associated reflexive and
transitive relation $\leq$ satisfies relatively few properties,
since adding constraints does not lead to worsen the solution,
thus resulting in a non-monotonic framework~\cite{nic}.

\section{QoS Routing}
\label{sec:qosrouting}

With \emph{Constraint-Based Routing} (CBR) we refer to a class of
routing algorithms that base path selection decisions on a set of
requirements or constraints, in addition to destination criteria.
These constraints may be imposed by administrative policies (i.e.
\emph{policy routing}), or by QoS requirements (i.e. QoS routing,
as already cited in Sec.~\ref{sec:introduction}), and so they can
be classified in two classes with different characteristics. The
aim of CBR is to reduce the manual configuration and intervention
required for attaining traffic engineering
objectives~\cite{RFC3031}; for this reason, CBR enhances the
classical routing paradigm with special properties, such as being
resource reservation-aware and demand-driven.

The routing associated with administration decisions is referred
to as \emph{policy routing} (or policy-based routing), and it is
used to select paths that conform to administrative rules and
\emph{Service Level Agreements} (SLAs) stipulated among service
providers and clients. In this way, routing decisions can be based
not only on the destination location, but also on other factors
such as applications or protocols used, size of packets, identity
of the communicating entities, or in general, business related
decisions. Policy constraints can help improving the global
security of the network: constraints can be used to guarantee
agreed service provisioning and safety from malicious users
attempting to steal the resources not included in their contracts.

QoS routing attempts to simultaneously satisfy multiple QoS
requirements requested by real-time applications: the requirements
are usually expressed using metrics as, e.g. delay and bandwidth.
Policy routing (or policy-based routing) is instead used to select
paths that conform to imposed administrative rules. In this way,
routing decisions can be based not only on the destination
location, but also on factors such as used applications and
protocols, size of packets, or identity of both source and
destination end systems of the flow. Policy constraints can
improve the global security of network infrastructure and are able
to realize business related decisions.

Traditionally, QoS metrics can be organized into three distinct
classes, depending on how they are combined along a path: they can
be \emph{i) additive}, \emph{ii) multiplicative} or \emph{iii)
concave}~\cite{NP2}. They are defined as follows: with $n_1, n_2,
n_3 \dots, n_i, n_j$ representing network nodes, let $m(n_1 ,
n_2)$ be a metric value for the link connecting $n_1$  and $n_2)$.
For any path $P = (n , n_2 , \dots, n_i , n_j )$, the metric
corresponding is:

\begin{itemize}
\item \emph{Additive}, if $m(P) = m(n_1 ,n_2) + m(n_2 , n_3) + ... + m(n_i ,
n_j)$ The additive metric of a path is the sum of the metric for
all the links constituting the path. Some examples are delay,
jitter (the delay variation on a network path), cost and
hop-count.
\item \emph{Multiplicative}, if $m(P) = m(n_1 , n_2) \times m(n_2 , n_3) \dots \times m(n_i
, n_j)$ Multiplicative metric (equivalent to the additive one just
by taking the logarithm of all costs) of a path consists in the
multiplication of the metric values for all the links constituting
the path. Example is reliability or loss probability.
\item \emph{Concave}, if $m(P) = \max \slash \min \{m(n_1 , n_2), m(n_2 , n_3), ... , m(n_i
, n_j)\}$. The concave metric of a path is the maximum or the
minimum of the metric values over all the links in the path. The
classical example is bandwidth, meaning that the bandwidth of a
path is determined by the link with the minimum available
bandwidth, i.e. the bottleneck of the path. Other concave metrics
can be represented for example by packet buffers or CPU usage of
the routers along the path, or, however, something to be maximized
and depending on the ``weakest link''.
\end{itemize}

Even if usually the metric classes are introduced for paths, most
of times they can be suitable also for trees: consider, for
example, if we need to find a global cost of the tree by summing
up all the weights on the tree edges (i.e. additive), or if we
want to maximize the bandwidth of bottleneck link (i.e. concave).

Given a node generating packets, we can classify network data
delivery schemas into three main classes: \emph{i) unicast}, when
data is delivered from one sender to one specific recipient,
providing one-to-one delivery, \emph{ii) broadcast}, when data is
instead delivered to all hosts, providing one-to-all delivery, and
finally, \emph{iii) multicast}, when data is delivered to all the
selected hosts that have expressed interest; thus, this last
method provides one-to-many delivery. We will concentrate on
\emph{i)} and \emph{iii)}.


\subsection{Two NP-Complete Problems}\label{sec:npcomplete}When we use multiple QoS metrics, a typical
scenario involves resources that are independent and allowed to
take real or unbounded integer values~\cite{NP5}. For example, it
could be necessary to find a route with the objective of cost
minimization (i.e. a \emph{quantitative} constraint, optimizing a metric) and subject
to a path delay $\leq$ 40msec (i.e. a \emph{boolean} constraint, saying whether or not a route
is feasible)
at the same time, therefore we would have the set of constraints
$C = (delay \leq 40, \min (cost))$. In such scenarios, satisfying
two boolean constraints, or a boolean constraint and a
quantitative (optimization) constraint is NP-Complete~\cite{CBR}.
If all resources except one take bounded integer values, or if
resources are dependent, then the problems can be solved in
polynomial time~\cite{NP4}. Most of the proposed algorithms in
this area apply heuristics to reduce the complexity, as we will
see in Sec.~\ref{sec:unicastbg} and~\ref{sec:multicastbg}.

Unicast and multicast QoS routing can be reduced to two well-known
and more general problems: respectively, \emph{Multi-Constrained
Path} (MCP)~\cite{unicastbg2,qosnewsurvey} and \emph{Steiner Tree}
(ST)~\cite{Steiner87,qosnewsurvey} problems.
In MCP, the problem is to find a path from node $s$ to node $t$ in
a graph where each link is associated with $k$ non-negative
additive weights, while satisfying  a set of constraints $C$ on
these weights.

%
%
%

There may be multiple different paths in the graph $G(N,E)$ that
satisfy the same set of constraints. Such paths are said to be
feasible. 
However, often it might be desirable to retrieve an optimal path,
according to some criteria, and respecting also the bounds imposed
by the constraints. This more difficult problem is known as the
\emph{Multi-Constrained Optimal Path} (MCOP) problem. Clearly,
since the paths must be optimized according to some costs
criteria, MCOP intersects the Shortest Path problem.

%

The MCP problem is a NP-Complete problem. The authors
of~\cite{NP1} were the first to list the MCP problem with a number
of metrics $m = 2$ as being NP-complete, but they did not provide
a proof. Wang and Crowcroft have provided this proof for $m \geq
2$ in~\cite{NP2} and~\cite{NP3}, which basically consisted of
reducing the MCP problem for $m = 2$ to an instance of the
partition problem, a well-known NP-complete problem~\cite{NP1}.
However, simulations performed in (for
example)~\cite{heurqos1,heurqos2,CBR} show that QoS routing may be
practically tractable in some of the possible cases.

In the ST problem, given a set $S$ of vertices in a graph $G=(V,
E)$, a solution interconnects them by a graph of minimum weight,
where the weight is the sum of the weights of all edges. If $S =
V$, the ST problem reduces to the \emph{Minimum Spanning Tree}
(MST) problem\cite{Cormen}. 
ST has been extended to \emph{Constrained Steiner Tree} (CST), to
include constraints concerning the weights of the links; for
example, if we want that the sum of the metric values for each
path $p$ from the source $s$ to each leaf $s \in S$, is less than
a chosen limit $\Delta$. ST and CST are NP-Complete
problems~\cite{Steiner87} since the second can be reduced to the
first one.

As can be seen, the problems related to multicast inherit both the
difficulty of multiple constrained metrics, and the difficulty to
reach multiple end-nodes at the same time.


%


\subsection{Unicast Routing with QoS Extensions}\label{sec:unicastbg}

\begin{figure}
\centering
\includegraphics[scale=0.36]{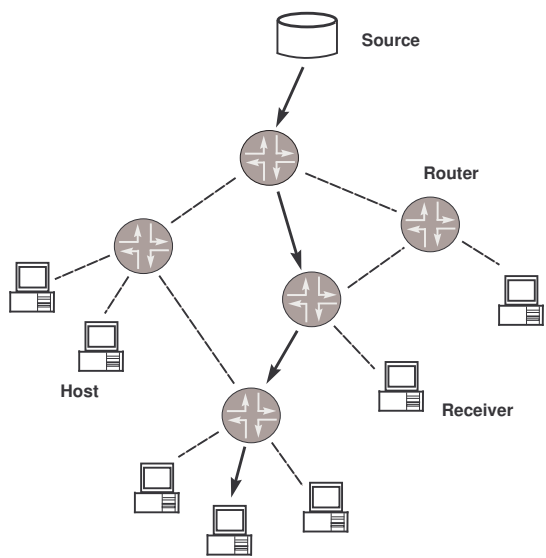} 
\caption{An example of a unicast distribution between the source
and the receiver. Oriented arcs highlight the path, while dashed
lines correspond to links not traversed by the flow.}
\label{figure:unicast}
\end{figure}

In Fig.~\ref{figure:unicast} we show an example of a unicast
communication between the \emph{source}, generating data, and the
only one \emph{receiver} (i.e. the destination of the
communication): the thick oriented lines highlight the direction
of the packet flow, while dashed lines correspond to links not
traversed.

Now we present some of the unicast QoS routing proposals, each of
them oriented at optimizing only a small subset of the possible
QoS metrics or using heuristics, since, as presented in
Sec.~\ref{sec:npcomplete}, the problem is in general NP-Complete.
For example, several solutions have been proposed for
bandwidth-bounded routing: an interesting approach proposed
in~\cite{unicastbg1} exploits the dependencies among resources,
e.g. available bandwidth, delay, and buffer space, to simplify the
problem; then, a modified Bellman-Ford algorithm can be used to
solve the problem. One approach to satisfy both bandwidth and
delay bounds is to first prune all links not satisfying the
bandwidth requirement. Dijkstra's shortest path algorithm is then
applied to find a feasible path, if any, satisfying the delay
requirements~\cite{NP2}. The problem of optimizing both the
bandwidth and the delay can be either solved as a \emph{widest
shortest path} problem or a \emph{shortest widest path} problem,
depending if the algorithm gives higher priority to selecting
paths with minimum hop counts (i.e. \emph{widest shortest path}),
or to selecting paths with maximum bandwidth (i.e. \emph{shortest
widest path})~\cite{NP2}. The objective of multi-constrained
routing is to simultaneously satisfy a set of constraints, as
described in~\cite{unicastbg1,unicastbg2}. In~\cite{unicastbg2} is
proposed a heuristic approach for the multi-constrained optimal
path problem (defined a \emph{H\_MCOP}), which optimizes a
non-linear function (for feasibility) and a primary function (for
optimality). There are also solutions for bandwidth and cost
bounded routing, which typically map the cost or the bandwidth to
a bounded integer value, and then solve the problem in polynomial
time using an extended version of Bellman-Ford or Dijkstra
algorithms~\cite{NP4}.

\subsection{Multicast Routing with QoS extensions}
\label{sec:multicastbg}

\begin{figure}
\centering
\includegraphics[scale=0.36]{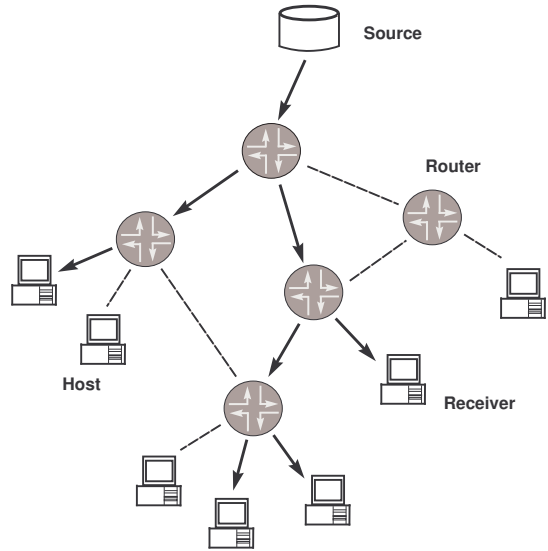} 
\caption{An example of a multicast tree built over a network:
oriented arcs highlight the tree (direction is down stream), while
dashed lines correspond to links not traversed by the flow.}
\label{figure:multicast}
\end{figure}

Multicast is an important bandwidth-conserving technology that
reduces traffic by simultaneously delivering a single stream of
information to multiple receivers (as shown in
Fig.~\ref{figure:multicast}). Therefore, while saving resources,
multicast is well suited to concurrently distribute contents on
behalf of applications asking for a certain timeliness of
delivery: thus, also multicast routing has naturally been extended
to guarantee QoS requirements~\cite{qosMulticast}. In its simplest
implementation, multicast can be provided using multiple unicast
transmissions (i.e. the source would be in charge to open them),
but with this solution, the same packet can traverse the same link
multiple times, thus increasing the network traffic. For this
reason, the network must provide this service natively, by
creating multicast (group) addresses and by letting the routers
duplicate the packet only when the distribution tree effectively
forks. In this way, the source node has to know only one global
address for all the destinations, and the network (i.e. the
routers) can optimally ``split'' the flow towards the receivers,
knowing also how to optimize traffic: the source node cannot have
this information.

A multicast address is also called a multicast group address, with
which the routers can locate and send packets to all the members
in the group. A group member is a host that expresses interest in
receiving packets sent to a specific group address. A group member
is also sometimes called a \emph{receiver} or a \emph{listener}. A
multicast source is a host that sends packets with the destination
address set to a multicast group. To deliver data only to
interested parties, routers in the network build a
\emph{multicast} (or \emph{distribution}) \emph{tree}
(Fig.~\ref{figure:multicast}). Each subnetwork that contains at
least one interested listener is a leaf of the tree. Where the
tree branches, routers replicate the data and send a single packet
down each branch. No link ever carries a duplicate flow of
packets, since packets are replicated in the network only at the
point where paths diverge, reducing the global traffic.

Multicast problem has been studied with several algorithms and
variants, such as \emph{Shortest-Path Tree} (SPT), MST, ST, CST
(see Sec.~\ref{sec:qosrouting}), and other miscellaneous
trees~\cite{qosMulticast}. Algorithms based on SPT (e.g. Dijkstra
or Bellman-Ford~\cite{Cormen}) aim to minimize the sum of the
weights on the links from the source to each receiver, and if all
the link cost one unit, the resulting tree is the least-hop one.

Multicast QoS routing is generally more complex than unicast QoS
routing, and for this reason less proposals have been elaborated
in this area~\cite{CBR,qosnewsurvey}. With respect to unicast, the
additional complexity stems from the need to support shared and
heterogeneous reservation styles (towards distinct group members)
and global admission control of the distribution flow. Some of the
approaches use a Steiner tree formulation~\cite{Steiner} or extend
existing algorithm to optimize the delay (i.e. MOSPF~\cite{MOSPF}
is the multicast version of the classical OSPF), while the
\emph{Delay Variation Multicast Algorithm} (DVMA)~\cite{DVMA}
computes a multicast tree with both bounded delay and bounded
jitter. Also, delay-bounded and cost-optimized multicast routing
can be formulated as a Steiner tree: an example approach is
\emph{QoS-aware Multicast Routing Protocol}~\cite{QMRP} (QMRP).
Other multicast QoS routing algorithms and related problems
(entailing stability, robustness and scalability) are presented
in~\cite{CBR}.

\section{Finding unicast QoS routes with SCLP Programs}\label{sec:unicastmodel}

In this Section we will show how to represent and solve unicast
QoS routing with SCLP. At the beginning the problem will be
treated only from the cost optimization view, i.e. as a SP
problem, while in the last part we propose an example on how to
add constraints on the path (i.e. solving the MCOP problem seen in
Sec.~\ref{sec:unicastbg}). Sec.~\ref{sec:sptoSCLP} translates SP
problems as SCLP programs, while in Sec.~\ref{sec:po} the same
model is extended for multi-criteria optimizations, thus featuring
vectors of costs on the edges, and not a single value.
Sec.~\ref{sec:mo} describes the case where each arc also stores
information about the {\em modality} to be used to traverse the
arc. At last, in Sec.~\ref{sec:addconstraints} we add constraints
on the QoS metrics, in order to fully obtain a model for
constrained paths.

\subsection{From SP Problems to SCLP Programs}\label{sec:sptoSCLP}

\begin{figure}
  \begin{center}
    \includegraphics[width=7cm]{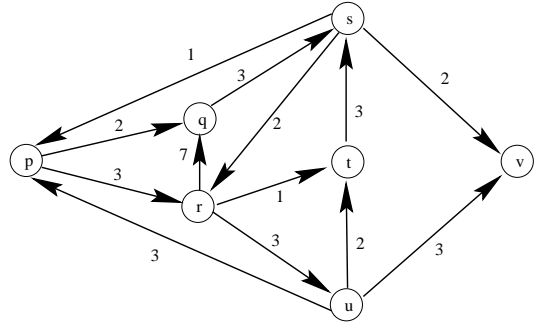}
    \caption{An SP problem.}
    \label{fig:spp}
  \end{center}
\end{figure}

We suppose to work with a graph $G = (N, E)$, where each oriented
arc $e \in E$ from node $p$ to node $q$ ($p,q \in N$) has
associated a label representing the cost of the arc from $p$ to
$q$, as the example in Fig~\ref{fig:spp}. This graph can be easily
used to represent a network, if nodes are associated to network
devices (routers and hosts) and arcs to network links. From any SP
problem we can build an SCLP program as follows.

For each arc we have two clauses: one describes the arc and the
other one its cost. More precisely, the head of the first clause
represents the starting node, and its body contains both the final
node and a predicate, say $c$, representing the cost of the arc.
Then, the second clause is a fact associating to predicate $c$ its
cost (which is a semiring element). Even if in this Section the
concept of cost is quite general, we recall that with this fact we
represent the QoS metric values on the arc (see
Sec.~\ref{sec:qosrouting}). For example, if we consider the arc
from $p$ to $q$ with cost $2$, we have the clause\\
{\tt p :- $c_{pq}$, q.}\\
and the fact \\
{\tt $c_{pq}$ :- 2.}\\
Finally, we must code that we want $v$ to be the final node of the
path. This is done by adding a clause of the form {\tt v :- 0.}
Note also that any node can be required to be the final one, not
just those nodes without outgoing arcs (like $v$ is in this
example). The whole program corresponding to the SP problem in
Fig.~\ref{fig:spp} can be seen in Table~\ref{tab:SP}.

\begin{table}
\caption{The SCLP program representing the SP problem in
Fig.~\ref{fig:spp}.} \label{tab:SP} \hrule \vskip3pt
\begin{tabbing}
{\tt $c_{pq}$} \= {\tt  :- $c_{pq}$, q.} blablabla bla blablabla
\= {\tt $c_{pq}$} \= \kill
{\tt p} \> {\tt :- $c_{pq}$, q.}  \> {\tt $c_{pq}$} \> {\tt :- 2.}\\
{\tt p} \> {\tt :- $c_{pr}$, r.} \> {\tt $c_{pr}$} \> {\tt :- 3.}\\
{\tt q} \> {\tt :- $c_{qs}$, s.} \> {\tt $c_{qs}$} \> {\tt :- 3.}\\
{\tt r} \> {\tt :- $c_{rq}$, q.} \> {\tt $c_{rq}$} \> {\tt :- 7.}\\
{\tt r} \> {\tt :- $c_{rt}$, t.} \> {\tt $c_{rt}$} \> {\tt :- 1.}\\
{\tt r} \> {\tt :- $c_{ru}$, u.} \> {\tt $c_{ru}$} \> {\tt :- 3.}\\
{\tt s} \> {\tt :- $c_{sp}$, p.} \> {\tt $c_{sp}$} \> {\tt :- 1.}\\
{\tt s} \> {\tt :- $c_{sr}$, r.} \> {\tt $c_{sr}$} \> {\tt :- 2.}\\
{\tt s} \> {\tt :- $c_{sv}$, v.} \> {\tt $c_{sv}$} \> {\tt :- 2.}\\
{\tt t} \> {\tt :- $c_{ts}$, s.} \> {\tt $c_{ts}$} \> {\tt :- 3.}\\
{\tt u} \> {\tt :- $c_{up}$, p.} \> {\tt $c_{up}$} \> {\tt :- 3.}\\
{\tt u} \> {\tt :- $c_{ut}$, t.} \> {\tt $c_{ut}$} \> {\tt :- 2.}\\
{\tt u} \> {\tt :- $c_{uv}$, v.} \> {\tt $c_{uv}$} \> {\tt :- 3.}\\
{\tt v} \> {\tt :- 0.}  \> \>
\end{tabbing}
\vskip1pt \hrule
\end{table}

To represent the classical version of SP problems, we consider
SCLP programs over the semiring $S = \langle \mathbb{N},
\mbox{min}, +, +\infty, 0 \rangle$, which is an appropriated
framework to represent constraint problems where one wants to
minimize the sum of the costs of the solutions. For example, we
can imagine that the cost on the arcs represents to us the average
delay experienced on the related link (measured in tens
milli-seconds). To compute a solution of the SP problem it is
enough to perform a query in the SCLP framework; for example, if
we want to compute the cost of the path from $r$ to $v$ we have to
perform the query {\tt :- r}. For this query, we obtain the value
$6$, that represents the cost of the best path(s) from $r$ to $v$,
optimizing in this way the total delay experienced on the route
from $r$ to $v$. Clearly, different semirings can be chosen to
represent the  composition properties of the different metrics, as
we will see better in Sec.~\ref{sec:po} by proposing bandwidth as
the second metric describing the link costs.

Notice that to represent classical SP problems in SCLP, we do not
need any variable. Thus the resulting program is propositional.
However, this program, while giving us the cost of the shortest
paths, does not give us any information about the arcs which form
such paths. This information could be obtained by providing each
predicate with an argument, which represents the arc chosen at
each step.

\begin{figure}
  \begin{center}
    \includegraphics[width=7cm]{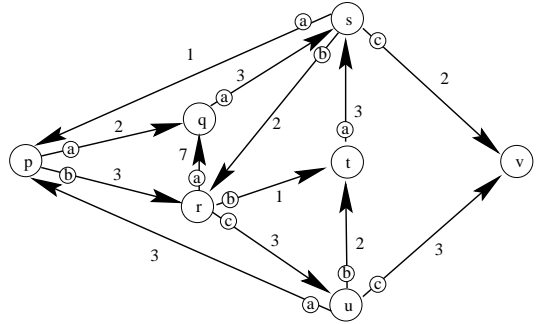}
    \caption{An SP problem with labeled arcs.}
    \label{fig:spplab}
  \end{center}
\end{figure}

Figure~\ref{fig:spplab} shows the same SP problem of
Fig.~\ref{fig:spp} where the arcs outgoing each node have been
labeled with different labels to distinguish them. Such labels can
then be coded into the corresponding SCLP program to ``remember''
the arcs traversed during the path corresponding to a solution.
For example, clause\\
{\tt p :- $c_{pq}$, q.}\\
would be rewritten as \\
{\tt p(a) :- $c_{pq}$, q(X).}\\
Here constant $a$ represents one of the arcs going out of $p$: the
one which goes to $q$. If all clauses are rewritten similarly,
then the answer to a goal like {\tt :- r(X)} will be both a
semiring value (in our case $6$) and a substitution for $X$. This
substitution will identify the first arc of a shortest path from
$r$ to $v$. For example, if we have $X=b$, it means that the first
arc is the one that goes from $r$ to $t$. To find a complete
shortest path, we just need to compare the semiring values
associated with each instantiated goal, starting from $r$ and
following the path. For example, in our case (of the goal $\exists
X. r(X)$) we have that the answer to the goal will be $X=c$ with
semiring value $6$. Thus we know that a shortest path from $r$ to
$v$ can start with the arc from $r$ to $u$. To find the following
arc of this path, we compare the semiring values of $u(a)$,
$u(b)$, and $u(c)$. The result is that $u(c)$ has the smallest
value, which is $3$. Thus the second arc of the shortest path we
are constructing is the one from $u$ to $v$. The path is now
finished because we reached $v$ which is our final destination.

Notice that a shortest path could be found even if variables are
not allowed in the program, but more work is needed. In fact,
instead of comparing different instantiations of a predicate, we
need to compare the values associated with the predicates that
represent nodes reachable by alternative arcs starting from a
certain node, and sum them to the cost of such arcs. For example,
instead of comparing the values of $p(a)$ and $p(b)$
(Fig.~\ref{fig:spplab}), we have to compare the values of $q + 2$
and of $r + 3$ (Fig.~\ref{fig:spp}).

A third alternative to compute a shortest path, and not only its
cost,
is to use lists: by replacing each clause of the form\\
{\tt p :- c$_{xy}$, q.}\\
with the clause\\
{\tt p([a|T]) :- c$_{xy}$, q(T).}\\
during the computation we also build the list containing all arcs
which constitute the corresponding path. Thus, by giving the goal
{\tt :- p(L).}, we would get both the cost of a shortest path and
also the shortest path itself, represented by the list $L$.

An alternative representation, probably more familiar for CLP
users,
of SP problems in SCLP is one where there are facts of the form\\
{\tt c(p,q) :- 2.}\\
\vdots\\
{\tt c(u,v) :- 3.}\\
to model the graph, and the two clauses\\
{\tt path(X,Y) :- c(X,Y).}\\
{\tt path(X,Y) :- c(X,Z), path(Z,Y).}\\
to model paths of length one or more. In this representation the
goal to be given to find the cost of the shortest path from $p$ to
$v$ is {\tt :- path(p,v).} This representation is obviously more
compact than the one in Table~\ref{tab:SP}, and has equivalent
results and properties. However, in next Sections we will continue
using the simpler representation, used in Table~\ref{tab:SP},
where all the clauses have at most one predicate in the body. The
possibility of representing SP problems with SCLP programs
containing only such a kind of clauses is important, since it will
allow us to use efficient algorithms to compute the semantics of
such programs (see~\cite{sclpjheu} for more details).

\subsection{Partially-Ordered SP Problems} \label{sec:po}

\begin{figure}
  \centering
    \includegraphics[width=7cm]{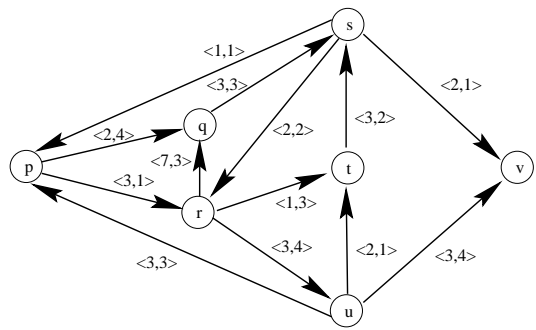}
    \caption{A multi-criteria SP problem.}
    \label{fig:multispp}
\end{figure}

Sometimes, the costs of the arcs are not elements of a totally
ordered set. A typical example is obtained when we consider
multi-criteria SP problems. Consider for example the
multi-criteria SP problem shown in Fig.~\ref{fig:multispp}: each
arc has associated a pair that  represent the weight of the arc in
terms of {\em cost} of use and average {\em delay} (i.e. two
possible QoS metrics); thus, the values are in the $\langle cost,
delay \rangle$ form. Given any node $p$, we want to find a path
from $p$ to $v$ (if it exists) that minimizes both criteria. In
this example, there may be cases in which the labels of two arcs
are not compatible, like $\langle 5, 20 \rangle$ and $\langle 7,
15 \rangle$, since the cost is better in the first pair, while the
delay is lower in the second one. In general, when we have a
partially ordered set of costs, it may be possible to have several
paths, all of which are not {\em dominated} by others, but which
have different incomparable costs (see also
Sec.~\ref{sec:partiallyref}).

We can translate this SP problem in Fig.~\ref{fig:multispp} into
the corresponding SCLP program in Table~\ref{tab:multispp}. This
program works over the semiring
$$\langle \mathbb{N}^2, \mbox{min'}, +', \langle +\infty, +\infty \rangle,
\langle 0, 0 \rangle \rangle,$$ where $min'$ and $+'$ are
classical $min$ and $+$, suitably extended to pairs. In practice,
this semiring is obtained by putting together, via the Cartesian
product, two instances of the semiring $\langle \mathbb{N},
\mbox{min}, +, +\infty, 0 \rangle$ (we recall that the Cartesian
product of two c-semirings is a c-semiring as well~\cite{jacm97}).
One of the two instances is used to deal with the cost criteria,
the other one is for the delay criteria. By working on the
combined semiring, we can deal with both criteria simultaneously:
the partial order will tell us when a $\langle cost, delay
\rangle$ pair is preferable to another one, and also when they are
not comparable.

\begin{table}
\caption{The SCLP program representing the multi-criteria SP
problem in Fig.~\ref{fig:multispp}.} \label{tab:multispp} \hrule
\vskip3pt
\begin{tabbing}
{\tt $c_{pq}$} \= {\tt  :- $c_{pq}$, q.} blablabla bla blablabla
\= {\tt $c_{pq}$} \= \kill
{\tt p} \> {\tt :- $c_{pq}$, q.} \> {\tt $c_{pq}$} \> {\tt :- < 2,4 >.}\\
{\tt p} \> {\tt :- $c_{pr}$, r.} \> {\tt $c_{pr}$} \> {\tt :- < 3,1 >.}\\
{\tt q} \> {\tt :- $c_{qs}$, s.} \> {\tt $c_{qs}$} \> {\tt :- < 3,3 >.}\\
{\tt r} \> {\tt :- $c_{rq}$, q.} \> {\tt $c_{rq}$} \> {\tt :- < 7,3 >.}\\
{\tt r} \> {\tt :- $c_{rt}$, t.} \> {\tt $c_{rt}$} \> {\tt :- < 1,3 >.}\\
{\tt r} \> {\tt :- $c_{ru}$, u.} \> {\tt $c_{ru}$} \> {\tt :- < 3,4 >.}\\
{\tt s} \> {\tt :- $c_{sp}$, p.} \> {\tt $c_{sp}$} \> {\tt :- < 1,1 >.}\\
{\tt s} \> {\tt :- $c_{sr}$, r.} \> {\tt $c_{sr}$} \> {\tt :- < 2,2 >.}\\
{\tt s} \> {\tt :- $c_{sv}$, v.} \> {\tt $c_{sv}$} \> {\tt :- < 2,1 >.}\\
{\tt t} \> {\tt :- $c_{ts}$, s.} \> {\tt $c_{ts}$} \> {\tt :- < 3,2 >.}\\
{\tt u} \> {\tt :- $c_{up}$, p.} \> {\tt $c_{up}$} \> {\tt :- < 3,3 >.}\\
{\tt u} \> {\tt :- $c_{ut}$, t.} \> {\tt $c_{ut}$} \> {\tt :- < 2,1 >.}\\
{\tt u} \> {\tt :- $c_{uv}$, v.} \> {\tt $c_{uv}$} \> {\tt :- < 3,4 >.}\\
{\tt v} \> {\tt :- < 0,0 >.} \> \>
\end{tabbing}
\vskip1pt \hrule
\end{table}

To give an idea of another practical application of
partially-ordered SP problems, just think of network routing
problems where we need to optimize according to the following
criteria: minimize the delay, minimize the cost, minimize the
number of arcs traversed, and maximize the bandwidth. The first
three criteria correspond to the same semiring, which is $\langle
\mathbb{N}, min, +, +\infty, 0 \rangle$, while the fourth criteria
can be characterized by the semiring $\langle \mathcal{B}, max,
min, 0, +\infty \rangle$, where $\mathcal{B}$ is the set of the
possible bandwidth values (in Sec.~\ref{sec:hypertranslation} we
will better investigate these semirings). In this example, we have
to work on a semiring which is obtained by vectorizing all these
four semirings. Each of the semirings is totally ordered but the
resulting semiring, whose elements are four-tuples, is partially
ordered.

\subsection{Modality-based SP Problems} \label{sec:mo}

Until now we have considered situations in which an arc is labeled
by its cost, be it one element or a tuple of elements as in the
multi-criteria case. However, sometimes it may be useful to
associate with each arc also information about the {\em modality}
to be used to traverse the arc.

For example, interpreting the arcs of a graph as links between
cities, we may want to model the fact that we can cover such an
arc by {\em car}, or by {\em train}, or by {\em plane}. Another
example of a modality could be the {\em time of the day} in which
we cover the arc, like {\em morning}, {\em afternoon}, and {\em
night}. One more example, this time strictly related to topic of
this paper, could be represented by the modalities associated with
the network link, e.g. {\em wired}, {\em wireless} or {\em VPN},
if there is the opportunity to establish a \emph{Virtual Private
Network} on it. Therefore the modalities can be used to manage
policies for the routing (i.e. for policy routing). In all these
examples, the cost of an arc may depend on its modality.

An important thing to notice is that a path could be made of arcs
which not necessarily are all covered with the same modality. For
example, the network connection between two distant buildings of
the same company can be made of many hops, some of which are
covered with the wireless modality and others with wired one.
Moreover, it can be that different arcs have different sets of
modalities. For example, from node $n_0$ to node $n_1$ we can use
both the wired or wireless connection, and from node $n_1$ to node
$n_2$ we can use only a VPN. Thus modalities cannot be simply
treated by selecting a subset of arcs (all those with the same
modality).

\begin{figure}
  \centering
    \includegraphics[width=7cm]{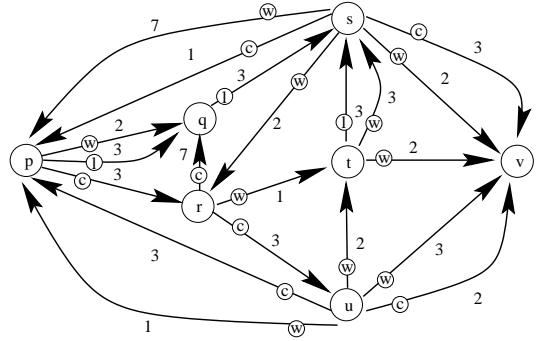}
    \caption{An SP problem with modalities.}
    \label{fig:mezzitrasp}
\end{figure}

An example of an SP problem with three modalities representing a
network with cryptographic service on the links (\emph{c}) (both
wired or wireless), wired/no-crypt (\emph{w}), and
wireless/no-crypt (\emph{l}) can be seen in
Fig.~\ref{fig:mezzitrasp}. Here the problem is to find a shortest
path from any node to $v$ (our final destination), and to know
both its delay and also the modalities of its arcs. This SP
problem can be modeled via the SCLP program in
Table~\ref{tab:mezzitrasp}. In this program, the variables
represent the modalities. If we ask the query {\tt :-p(c).}, it
means that we want to know the smallest delay for a route from $p$
to $v$ using the links with the cryptographic service. The result
of this query in our example is {\tt $p(c)=8$} (using the path
$p-r-u-v$).

\begin{table}
\caption{The SCLP program representing the SP problem with
modalities in Fig.~\ref{fig:mezzitrasp}.} \label{tab:mezzitrasp}
\hrule \vskip3pt
\begin{tabbing}
{\tt $c_{pq}$(X)} \= {\tt  :- $c_{pq}$, q.} blablabla bla
blablabla \= {\tt $c_{pq}$(X)} \= \kill
{\tt p(X)} \> {\tt :- $c_{pq}$(X), q(X).} \> {\tt $c_{pq}$(w)} \> {\tt :- 2.}\\
{\tt p(X)} \> {\tt :- $c_{pr}$(X), r(X).} \> {\tt $c_{pq}$(l)} \> {\tt :- 3.}\\
{\tt q(X)} \> {\tt :- $c_{qs}$(X), s(X).} \> {\tt $c_{pr}$(c)} \> {\tt :- 3.}\\
{\tt r(X)} \> {\tt :- $c_{rq}$(X), q(X).} \> {\tt $c_{qs}$(l)} \> {\tt :- 3.}\\
{\tt r(X)} \> {\tt :- $c_{rt}$(X), t(X).} \> {\tt $c_{rq}$(c)} \> {\tt :- 7.}\\
{\tt r(X)} \> {\tt :- $c_{ru}$(X), u(X).} \> {\tt $c_{rt}$(w)} \> {\tt :- 1.}\\
{\tt s(X)} \> {\tt :- $c_{sp}$(X), p(X).} \> {\tt $c_{ru}$(c)} \> {\tt :- 3.}\\
{\tt s(X)} \> {\tt :- $c_{sr}$(X), r(X).} \> {\tt $c_{sp}$(c)} \> {\tt :- 1.}\\
{\tt s(X)} \> {\tt :- $c_{sv}$(X), v(X).} \> {\tt $c_{sp}$(w)} \> {\tt :- 7.}\\
{\tt t(X)} \> {\tt :- $c_{ts}$(X), s(X).} \> {\tt $c_{sr}$(w)} \> {\tt :- 2.}\\
{\tt u(X)} \> {\tt :- $c_{up}$(X), p(X).} \> {\tt $c_{sv}$(w)} \> {\tt :- 2.}\\
{\tt u(X)} \> {\tt :- $c_{ut}$(X), t(X).} \> {\tt $c_{sv}$(c)} \> {\tt :- 3.}\\
{\tt u(X)} \> {\tt :- $c_{uv}$(X), v(X).} \> {\tt $c_{ts}$(l)} \> {\tt :- 3.}\\
{\tt v(X)} \> {\tt :- 0.} \> {\tt $c_{ts}$(w)} \> {\tt :- 3.}\\
\> \> {\tt $c_{up}$(c)} \> {\tt :- 3.}\\
\> \> {\tt $c_{up}$(w)} \> {\tt :- 1.}\\
\> \> {\tt $c_{ut}$(w)} \> {\tt :- 2.}\\
\> \>  {\tt $c_{uv}$(w)} \> {\tt :- 3.}\\
\> \> {\tt $c_{uv}$(c)} \> {\tt :- 2.}
\end{tabbing}
\vskip1pt \hrule
\end{table}

Notice that the formulation shown in Fig.~\ref{tab:mezzitrasp}
puts some possibly undesired constraints on the shortest path to
be found. In fact, by using the same variable in all the
predicates of a rule, we make sure that the same modality (in our
case the same transport mean) is used throughout the whole path.
If instead we want to allow different modalities in different arcs
of the path, then we just need to change the rules by putting a
new variable
on the last predicate of each rule. For example, the rule in Tab.~\ref{tab:mezzitrasp}\\
{\tt p(X) :- $c_{pq}$(X), q(X).}\\
would become \\
{\tt p(X) :- $c_{pq}$(X), q(Y).}\\
Now we can use a modality for the arc from $p$ to $q$, and another
one for the next arc. In this new program, asking the query {\tt
:-p(c).} means that we want to know the smallest delay for a trip
from $p$
to $v$ using the cryptographic service in the first arc. 

The same methods used in the previous Sections to find a shortest
path, or a non-dominated path in the case of a partial order, can
be used in this kind of SCLP programs as well. Thus we can put
additional variables in the predicates to represents alternative
arcs outgoing the corresponding nodes, and we can shift to the
semiring containing sets of costs to find a non-dominated path.
In particular, a clause like \\
{\tt p(X) :- $c_{pq}$(X), q(Y).}\\
would be rewritten as \\
{\tt p(X,a) :- $c_{pq}$(X), q(Y,Z).}

\subsection{Adding constraints to SP problems}
\label{sec:addconstraints} As seen in Sec.~\ref{sec:npcomplete} a
MCOP is much more difficult to solve than a SP problem, that is
NP-Complete. So far we considered only variants of SP problems
(partially-ordered or modality-based), but our aim is to provide a
complete model for the unicast QoS routing. Thus, besides
achieving cost optimization, we need also to consider constraints
on the QoS metrics.

In our example we consider again the multi-criteria graph in
Fig.~\ref{fig:multispp}: each arc has associated a pair that can
represent the weight of the arc in terms of {\em cost} of use and
average {\em delay}. However, in this case our goal is to minimize
the cost and to guarantee a final average delay  less than or
equal to $8$ ($80msec$), thus we want to add the boolean constraint $delay
\leq 8$.

We chose to represent constrained paths with a program in
\textit{CIAO Prolog}~\cite{CiaoManual}, a system that offers a
complete Prolog system supporting ISO-Prolog, but, at the same
time its modular design allows both restricting and extending the
basic language. 
CIAO Prolog has also a fuzzy extension, but since it does not
completely conform to the semantic of SCLP defined
in~\cite{ijcai97} (due to interpolation in the interval of the
fuzzy set), we decided to use the CIAO operators among constraints
(as $<$ and $\leq$), and to model the $\times$ operator of the
c-semiring with them. For this reason, we inserted the cost of the
edges in the head of the clauses, differently from SCLP clauses
which have the cost in the body of the clause. Similar reification processes
have been already accomplished also in other works~\cite{reification}.

In Tab.~\ref{tab:constrainedpath} is shown the CIAO program that
represents the graph in Fig.~\ref{fig:multispp}: here the edges
(i.e. all the \emph{Edges} facts in
Table~\ref{tab:constrainedpath}) are in the form:

$$edge(Source\_Node, Destination\_Node, [Link\_Cost, Link\_Delay])$$

\begin{table}
\caption{The CIAO program representing all the paths of
Fig.~\ref{fig:multispp}, with $delay \leq 8$} 
 \label{tab:constrainedpath}
\centering
\includegraphics[scale=0.63]{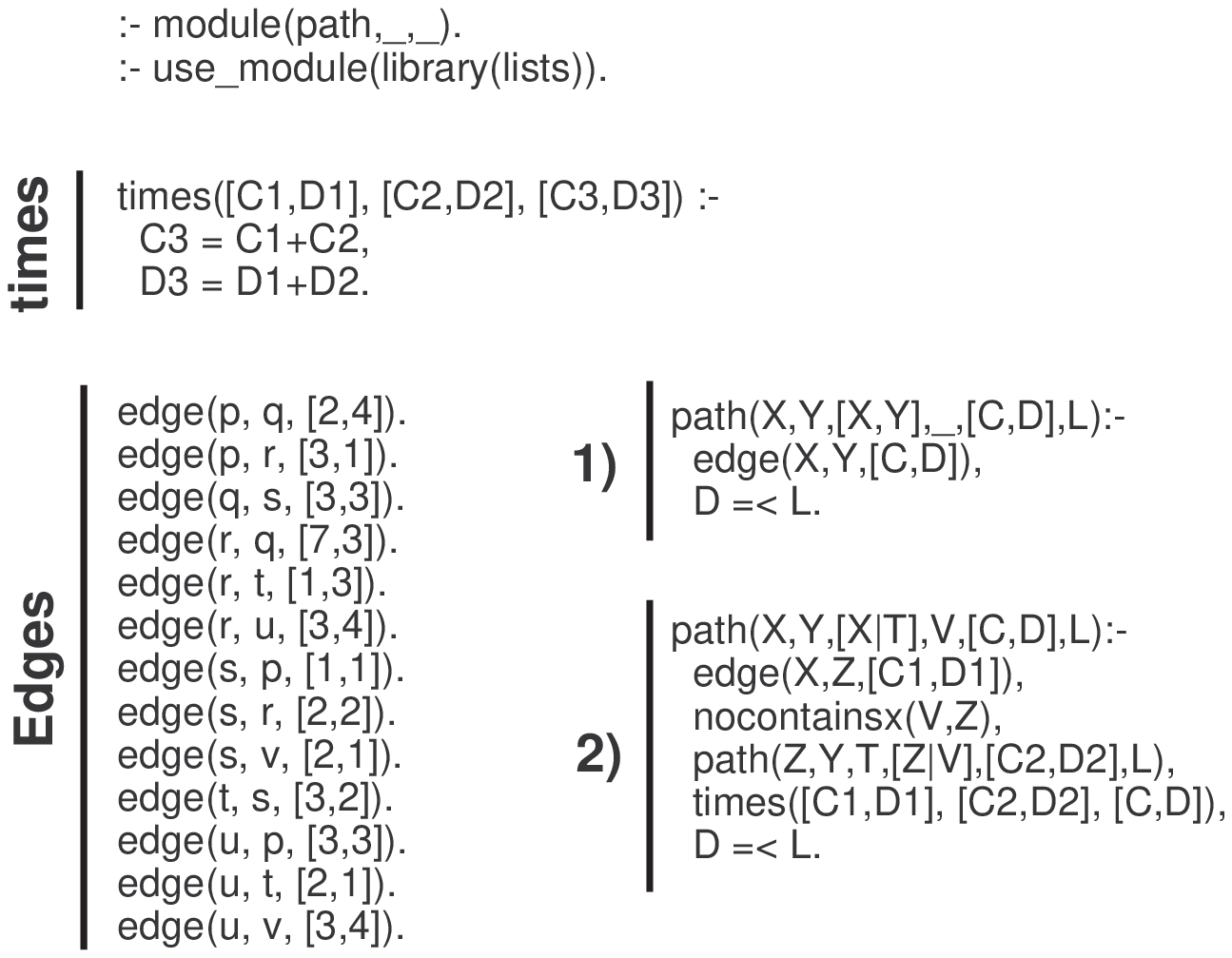}
\end{table}

Moreover, we can see the two clauses that describe the structure
of paths: \emph{Rule 1} and \emph{Rule 2} respectively represent
the base (or termination) case, where a path is simply an edge,
and the recursive case, needed to add one edge to the path. To
avoid infinite recursion, and thus the program crashing, we need
to deal with graph loops by considering the list of the already
visited nodes, in order to prevent the search from visiting them
twice. Moreover, we inserted a variable in the head of the
\emph{path} clauses to remember, at the end, all the visited nodes
of the path: this list will store the nodes following the correct
ordering of the visit. Finally, the last variable of the clause
head is used to retrieve only the paths with a total delay equal
or less than the passed value. Thus, the \emph{path} clause-heads
are in the form:

$$path(Source\_Node, Destination\_Node, Path\_Nodes,
Already\_Visisted\_Nodes,$$\vspace{-0.7cm}
$$[Path\_Cost, Path\_Delay], Path\_Max\_Delay)$$

The \emph{Aggregator} clause mimics the $\times$ operation of the
semiring (i.e. $+$ extended to pairs, as in Sec.~\ref{sec:po}),
and therefore it composes the global costs of the edges together,
edge costs with costs, and edge delays with delays.

All the paths with a $delay \leq 8$, and the relative query
$path(p,v,P,[p],[C, D],8)$ are shown in
Fig.~\ref{fig:outputprogram}. The $p$ source node of the path,
must be included in the list of the visited nodes from the
beginning. Figure~\ref{fig:outputprogram} corresponds to the
output of the CIAO program in Tab.~\ref{tab:constrainedpath}, and
for each of the three found paths it shows the variable $P$, which
stores the sequence of the nodes in the path, and the $C$ - $D$
pair, which corresponds to the  total cost of the path in terms of
$\langle cost, delay \rangle$.

We remark the expressivity of the framework, since boolean constraints
can be easily added to the query instead of being directly hard coded in the program.
For example, with a query like $path(p,v,P,[p],[C, D]), D<8$ returns all the paths
with a delay value less than $8$.

\begin{figure}
\centering
\includegraphics[scale=0.63]{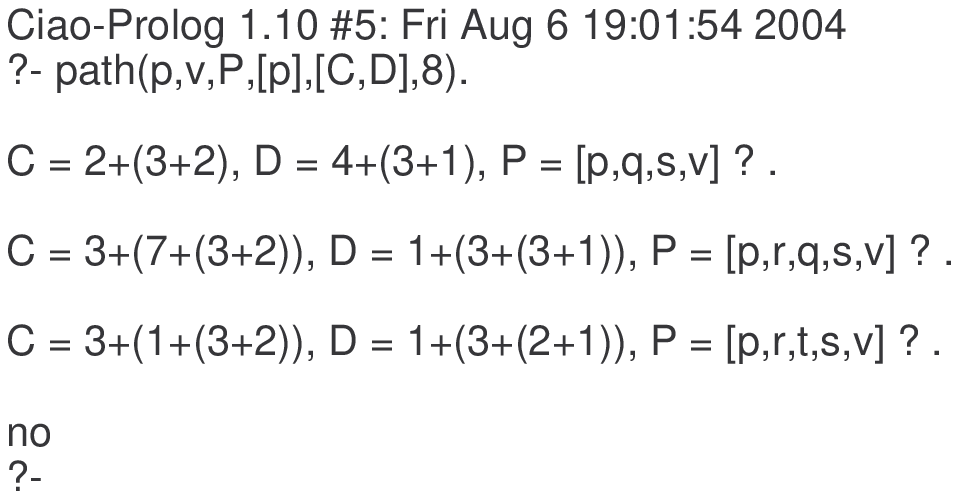}
\caption{The CIAO output for the program in
Tab.~\ref{tab:constrainedpath}: three paths are found with $delay
\leq 8$.} \label{fig:outputprogram}
\end{figure}

\section{Extending the Model to Deal with Multicast QoS Routing}
\label{sec:multicastmodel}

Now we extend the framework given in Sec.~\ref{sec:unicastmodel}
in order to manage also the multicast delivery schema. The first
step is represented by the use of hypergraphs instead of simple
graphs, since we need a method to connect one node to multiple
destinations at the same time (i.e. when the same packet must be
routed on different links). Section~\ref{sec:hypertranslation}
presents a possible transformation procedure from networks to
\emph{and-or} graphs, showing also how to find a cost for the
hyperarcs and related semirings. In Sec.~\ref{sec:SCLPandGraphs}
we describe the SCLP programs representing and solving the
multicast QoS routing. In Sec.~\ref{sec:modality-trees} we
associate modalities to hyperarcs, as we did in Sec.~\ref{sec:mo}
for paths.

\subsection{From networks to hypergraphs}
\label{sec:hypertranslation}
\begin{figure}
  \centering
  \includegraphics[scale=0.44]{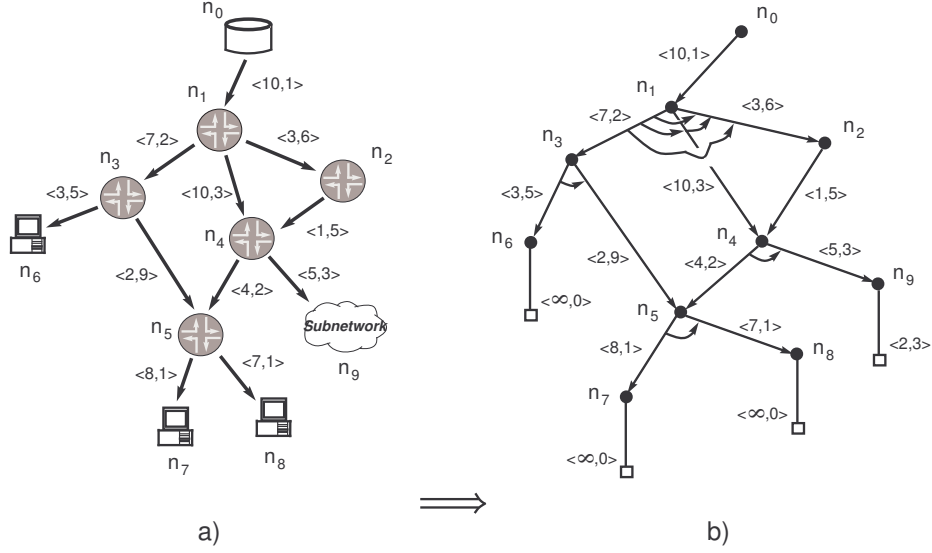} 
  \caption{A network example and the corresponding \emph{and-or} graph representation.}
  \label{translation}
\end{figure}

In this Section we explain a method to
translate the representation of a multicast network with QoS
requirements (Fig.~\ref{translation}\emph{a}) into a corresponding
weighted \emph{and-or} graph~\cite{mm78}
(Fig.~\ref{translation}\emph{b}). This procedure can be split in
three distinct steps, respectively focusing on the representation
of \emph{i)} network nodes, \emph{ii)} network links and
\emph{iii)} link costs in terms of QoS metrics.

An \emph{and-or} graph~\cite{mm78} is defined essentially as a
hypergraph. Namely, instead of arcs connecting pairs of nodes
there are hyperarcs connecting an $n$-tuple of nodes
($n=1,2,3,\ldots$). Hyperarcs are called {\em connectors} and they
must be considered as directed from their first node to all
others. Formally an \emph{and-or} graph is a pair
$G=(N,C)$, 
where $N$ is a set of {\em nodes} and $C$ is a set of connectors
\[C\subseteq N \times \bigcup_{i=0}^k N^i.\]

Note that the definition allows $0$-connectors, i.e. connectors
with one input and no output node. $0$-connectors are represented
as a line ending with a square (Fig.~\ref{translation}\emph{b}).
In the following of the explanation we will also use the concept
of \emph{and} tree~\cite{mm78}: given an \emph{and-or} graph $G$,
an \emph{and} tree $H$ is a {\em solution tree of $G$ with start
node $n_r$}, if there is a function $g$ mapping nodes of $H$ into
nodes of $G$ such that:
\begin{itemize}
\item the root of $H$ is mapped in $n_r$.
\item if ($n_{i_0}, n_{i_1}, \ldots, n_{i_k}$) is a connector of $H$,
  then ($g(n_{i_0}), g(n_{i_1}), \ldots, g(n_{i_k})$) is a connector of
  $G$.
\end{itemize}

In words, a solution tree of an \emph{and-or} graph is analogous
to a path of an ordinary graph: it can be obtained by selecting
exactly one outgoing connector for each node.

Each of the network nodes can be easily cast in the corresponding
\emph{and-or} graphs as a single graph node: thus, each node in
the graph can represent an interconnecting device (e.g. a router),
or a node acting as the source of a multicast communication
(injecting packets in the network), or, finally, a receiver
belonging to a multicast group and participating in the
communication. In Sec.~\ref{sec:SCLPandGraphs}, when we will look
for the best tree solution, the root of the best \emph{and} tree
will be mapped to the node representing the source of the
multicast communication; in the same way, receivers will be
modelled by the leaves of the resulting \emph{and} tree. When we
translate a receiver, we add an outgoing $0$-connector to model
the end-point of the communication, and whose cost will be
explained below. Suppose that $\{n_{0}, n_{1}, \dots , n_{9}\}$ in
Fig.~\ref{translation}\emph{a} are the identifiers of the network
nodes.

\begin{figure}
\centering
\includegraphics[scale=0.38]{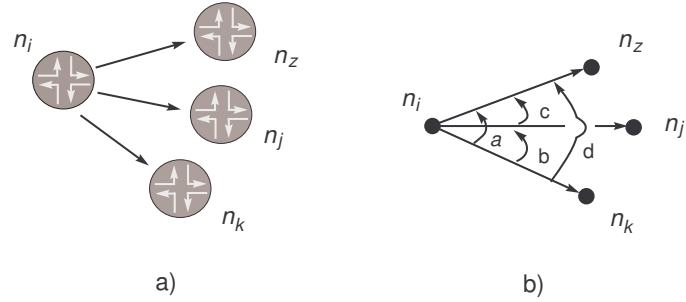}
\caption{\emph{a)} the f-star of  $n_{i}$ network-node and
\emph{b)} its representation with connectors.} \label{connector1}
\end{figure}

To model the links, we examine the forward star (\emph{f-star}) of
each node in the network (i.e. the set of arcs outgoing from a
node): we consider the links as oriented, since the cost of
sending packets from node $n_{i}$ to $n_{j}$ can be different from
the cost of sending from $n_{j}$ to $n_{i}$ (one non-oriented link
can be easily replaced by two oriented ones). Supposing that the
f-star of node $n_{i}$ includes the arcs $(n_{i}, n_{j})$,
$(n_{i}, n_{k})$ and $(n_{i}, n_{z})$, we translate this f-star by
constructing one connector directed from $n_{i}$ to each of the
subsets of destination nodes $\{j,k,z\}$ (Fig.~\ref{connector1}),
for a possible maximal number of $2^{|N|}-1$ subsets (where $|N|$
is the cardinality of the set of node in the graph), i.e.
excluding the emptyset; in Sec.~\ref{sec:complexity} we will see
how to minimize this exponential growth. Thus, all the resulting
connectors with $n_{i}$ as the input node are $(n_{i}, n_{j})$,
$(n_{i}, n_{k})$, $(n_{i}, n_{z})$, $(n_{i}, n_{k}, n_{j})$,
$(n_{i}, n_{k}, n_{z})$, $(n_{i}, n_{j}, n_{z})$ and $(n_{i},
n_{j}, n_{k}, n_{z})$. In the connectors tuple-ordering of the
nodes, the input node is at the first position and the output
nodes (when more than one) follow the orientation of the related
arrow in Fig.~\ref{connector1}.

To  simplify Fig.~\ref{connector1}\emph{b}, the arcs linking
directly two nodes represent 1-connectors $(n_{i}, n_{j})$,
$(n_{i}, n_{k})$ and $(n_{i}, n_{z})$, while curved oriented lines
represent $n$-connectors (with $n > 1$), where the set of their
output nodes corresponds to the output nodes of the traversed
arcs. With respect to $n_{i}$, in Fig.~\ref{connector1} we have a
curved line labelled with \emph{a} that corresponds to $(n_{i},
n_{k}, n_{j}, n_{z})$, \emph{b} to $(n_{i}, n_{k}, n_{j})$,
\emph{c} to $(n_{i}, n_{j}, n_{z})$, and, at last, \emph{d} to
$(n_{i}, n_{k}, n_{z})$. To have a clear figure, the network links
in Fig.~\ref{translation}\emph{a} are oriented ``towards'' the
receivers, thus we put only the corresponding connectors in
Fig.~\ref{translation}\emph{b}.

In the example we propose here, we are interested in QoS
link-state information concerning only bandwidth and cost.
Therefore, each link of the network can be labeled with a
2-dimensional cost, for example the pair $\langle 7, 3 \rangle$
tells us that the maximum bandwidth on that specific link is
70Mbps and the cost is 30{\euro}. In general, we could have a cost
expressed with a $n$-dimensional vector, where $n$ is the number
of metrics to be taken in account while computing the best
distribution tree. Since we want to maintain this link state
information even in the \emph{and-or} graph, we label the
corresponding connector with the same tuple of values
(Fig.~\ref{translation}).

In the case when a connector represent more than one network link
(i.e. a $n$-connector with $n \geq 2$), its cost is decided by
assembling the costs of the these links with the composition
operation $\circ$, which takes as many $n$-dimensional vectors as
operands, as the number of links represented by the connector.
Naturally, we can instantiate this operation for the particular
types of costs adopted to express QoS: for the example given in
this Section, the result of $\circ$ is the minimum bandwidth and
the highest cost (it could be also the sum of all the costs of the
links), ergo, the worst QoS metric values:
$$\circ (\langle b_1,c_1 \rangle , \langle b_2, c_2\rangle , \dots
, \langle b_n,c_n \rangle) \longrightarrow \langle \mbox{min}
(b_1, b_2, \dots , b_n) , \mbox{max} (c_1, c_2, \dots , c_n)
\rangle\ $$ The cost of the connector $(n_{1}, n_{3}, n_{4})$ in
Fig.~\ref{translation}\emph{b} will be $\langle 7,3 \rangle$,
since the costs of connectors $(n_{1}, n_{3})$ and $(n_{1},
n_{4})$ are respectively $\langle 7,2 \rangle$ and $\langle
10,3\rangle$:
$$\circ (\langle 7,2 \rangle , \langle 10,3\rangle) = \langle 7,3 \rangle$$
To simplify Fig.~\ref{translation}\emph{b}, we inserted only the
costs for the $1$-connectors, but the costs for the other
connectors can be easily computed with the $\circ$ operation, and
are all reported in Tab.~\ref{tab:and-or-graph}.

So far, we are able to translate an entire network with QoS
requirements in a corresponding \emph{and-or} weighted graph, but
still we need some algebraic framework to model our preferences
for the links to use in the best tree. For this reason, we use the
semiring structure (Sec.~\ref{sec:softCLP}). An exhaustive
explanation of the semiring framework approach for
shortest-distance problems is presented in~\cite{Mohri,tarjan}.

For example, if we are interested in maximizing the bandwidth of
the distribution tree, we can use the semiring $S_{Bandwidth} =
\langle {\cal B} \cup \{0 , +\infty\}, \mbox{max}, \mbox{min}, 0,
+\infty \rangle$; otherwise, we could be interested in minimizing
the global bandwidth with $\langle {\cal B} \cup \{0 , +\infty\},
\mbox{max}, \mbox{min}, +\infty, 0 \rangle$, if our intention is
to use an already busy link in order to preserve other unloaded
links for future use (i.e. for traffic engineering purposes). We
can use $S_{Money} = \langle \mathbb{N}, min, +, +\infty, 0
\rangle$ for the money cost, if we need to minimize the total cost
of the tree. Elements of $\cal B$ (i.e. the set of bandwidth
values) can be obtained by collecting information about the
network configuration, the current traffic state and technical
information about the links. Since the composition of c-semirings
is still a c-semiring~\cite{jacm97},
$$S_{Network} = \langle \langle {\cal B} \cup \{0, +\infty\},
\mathbb{N} \rangle, +', \times', \langle 0, +\infty \rangle,
\langle +\infty, 0 \rangle \rangle$$ where $+'$ and $\times'$
correspond to the vectorization of the $+$ and $\times$ operations
in the two c-semirings: given $b_{1}, b_{2} \in {\cal B} \cup \{0,
+\infty\}$ and $c_{1}, c_{2} \in \mathbb{N}$,
$$ \langle b_{1}, c_{1} \rangle +' \langle b_{2}, c_{2} \rangle = \langle
 \mbox{max} (b_{1}, b_{2}), \mbox{min} (c_{1},c_{2}) \rangle $$
$$\langle b_{1}, c_{1} \rangle \times' \langle b_{2}, c_{2} \rangle = \langle
 \mbox{min} (b_{1}, b_{2}), c_{1} + c_{2} \rangle$$

Clearly, the problem of finding best distribution tree is
multi-criteria, since both bandwidth and cost must be optimized.
We consider the criteria as independent among them, otherwise they
can be rephrased to a single criteria. Thus, the multidimensional
costs of the connectors are not elements of a totally ordered set,
and it may be possible to obtain several trees, all of which are
not {\em dominated} by others, but which have different
incomparable costs.

For each receiver node, the cost of its outgoing $0$-connector
will be always included in every tree reaching it. As a remind, a
$0$-connector has only one input node but no destination nodes. If
we consider a receiver as a plain node, we can set the cost as the
$1$ element of the adopted c-semiring ($1$ is the unit element for
$\times$), since the cost to reach this node is already completely
described by the other connectors of the tree branch ending in
this node: practically, we associate the highest possible QoS
values to this $0$-connector, in this case infinite bandwidth and
null cost. Otherwise we can imagine a receiver as a more complex
subnetwork (as the node $n_9$ in Fig.~\ref{translation}), and thus
we can set the cost of the $0$-connector as the cost needed to
finally reach a node in that subnetwork (as the cost $\langle 2,3
\rangle$ for the $0$-connector after node $n_9$ in
Fig.~\ref{translation}\emph{b}), in case we do not want, or
cannot, show the topology of the subnetwork, e.g. for security
reasons.

\subsection{\emph{And-or} graphs using SCLP} \label{sec:SCLPandGraphs}
In this Section, we represent an \emph{and-or} graph with a
program in SCLP. Using this framework, we can easily solve the
multi-criteria example concerning the multicast QoS network in
Fig.~\ref{translation}\textit{b}.

As already proposed in Sec.~\ref{sec:unicastmodel}, to represent
the connectors in SCLP we can write clauses like $c(n_i, [n_j,
n_k]) :- \langle 10, 3 \rangle$, stating that the graph has
connector from $n_i$ to nodes $n_j$ and $n_k$ with a bandwidth
cost of $100$Mbps and a cost of $30${\euro}. Other SCLP clauses
can properly describe the structure of the tree we desire to
search over the graph.

For the same reasons exposed in Sec.~\ref{sec:addconstraints}, we
choose to represent an \emph{and-or} graph with a program in
\textit{CIAO Prolog}~\cite{CiaoManual}. As an example, from the
weighted \emph{and-or} graph problem in
Fig.~\ref{translation}\emph{b} we can build the corresponding CIAO
program of Table~\ref{tab:and-or-graph} as follows. The set of
network edges (or $1$-connectors) is highlighted as \emph{Edges}
in Tab.~\ref{tab:and-or-graph}. Each fact has the structure
$$edge(source\_node, [dest\_nodes], [bandwidth, cost])$$
e.g. the fact $edge(n_0, [n_1], [10, 1])$ represents the
$1$-connector of the graph $(n_0,n_1)$ with bandwidth equal to
$100$Mbps and cost $10${\euro}. The \emph{Rules 1} in
Tab.~\ref{tab:and-or-graph} are used to compose the edges (i.e.
the $1$-connectors) together in order to find all the possible
$n$-connectors with $n \geq 1$, by aggregating the costs of
$1$-connectors with the $\circ$ composition operator, as described
in Sec.~\ref{sec:hypertranslation} (the lowest of the bandwidths
and the greatest of the costs of the composed $1$-connectors).
Therefore, with these clauses (in \emph{Rules 1}) we can
automatically generate the set of all the connectors outgoing from
the considered node (in Table~\ref{tab:and-or-graph},
\emph{nocontainsx} and \emph{insert\_last} are CIAO predicates
used to build a well-formed connector). The \emph{Leaves} in
Table~\ref{tab:and-or-graph} represent the $0$-connectors (a value
of $1000$ represents $\infty$ for bandwidth). The \emph{plus} and
\emph{times} rules in Table~\ref{tab:and-or-graph} respectively
mimic the $+$ and $\times$ operations of the semiring proposed in
Sec.~\ref{sec:hypertranslation}: $S_{Network} = \langle \langle
{\cal B} \cup \{0, +\infty\}, \mathbb{N} \rangle, +', \times',
\langle 0, +\infty \rangle, \langle +\infty, 0 \rangle \rangle$,
where $+'$ is equal to $\langle \mbox{max}, \mbox{min} \rangle$
and $\times'$ is equal to $\langle \mbox{min}, + \rangle$, as
defined in Sec.~\ref{sec:hypertranslation}. At last, the rules
$2$-$3$-$4$-$5$ of Tab.~\ref{tab:and-or-graph} describe the
structure of the routes we want to find over the graph. \emph{Rule
2} represents a route made of only one leaf node, \emph{Rule 3}
outlines a route made of a connector plus a list of sub-routes
with root nodes in the list of the destination nodes of the
connector, \emph{Rule 4} is the termination for \emph{Rule 5}, and
\emph{Rule 4} is needed to manage the junction of the disjoint
sub-routes with roots in the list $[X|Xs]$; clearly, when the list
$[X|Xs]$ of destination nodes contains more than one node, it
means we are looking for a multicast route. When we compose
connectors or trees (\emph{Rule 2} and \emph{Rule 5}), we use the
\emph{times} rule to compose their costs together. In \emph{Rule
5}, \emph{append} is a CIAO predicate used to join together the
lists of destination nodes, when the query asks for a multicast
route. At last, the \emph{route} predicate in
Sec.~\ref{sec:hypertranslation} collects all the results for the
query and finally returns the solution chosen with the help of the
\emph{plus} predicate.

Notice that the $\circ$ operator describes in
Sec.~\ref{sec:hypertranslation} is modeled with Prolog clauses
inside \emph{Rule 5}, when composing multiple $1$-connectors
connectors.

Notice also that the complexity of \emph{append} predicates in Tab.~\ref{tab:and-or-graph} can be reduced by
using \emph{difference} lists instead. However, see Sec.~\ref{sec:complexity} for complexity considerations.

To make the program in Tab.~\ref{tab:and-or-graph} as readable as
possible, we omitted two predicates: the \emph{sort} predicate,
needed to order the elements inside the list of destination-nodes
of connectors and trees (otherwise, the query $route(n_{0},
[n_{6},n_{7},n_{8},n_{9}], [B, C])$ and $route(n_{0},
[n_{9},n_{7},n_{8},n_{6}], [B, C])$ would produce different
results), and the \emph{intersection} predicate to check that
multiple occurrences of the same node do not appear in the same
list of destination nodes, if reachable with different connectors
(otherwise, for example, the tree $n_0, [n_7, n_7, n8, n9]$ would
be a valid result).

\begin{table}
\caption{The CIAO program representing the best result tree over
the weighted \emph{and-or} graph problem in
Fig.~\ref{translation}\emph{b}.} \hrule \vskip3pt
 \label{tab:and-or-graph}
\centering
\includegraphics[scale=0.63]{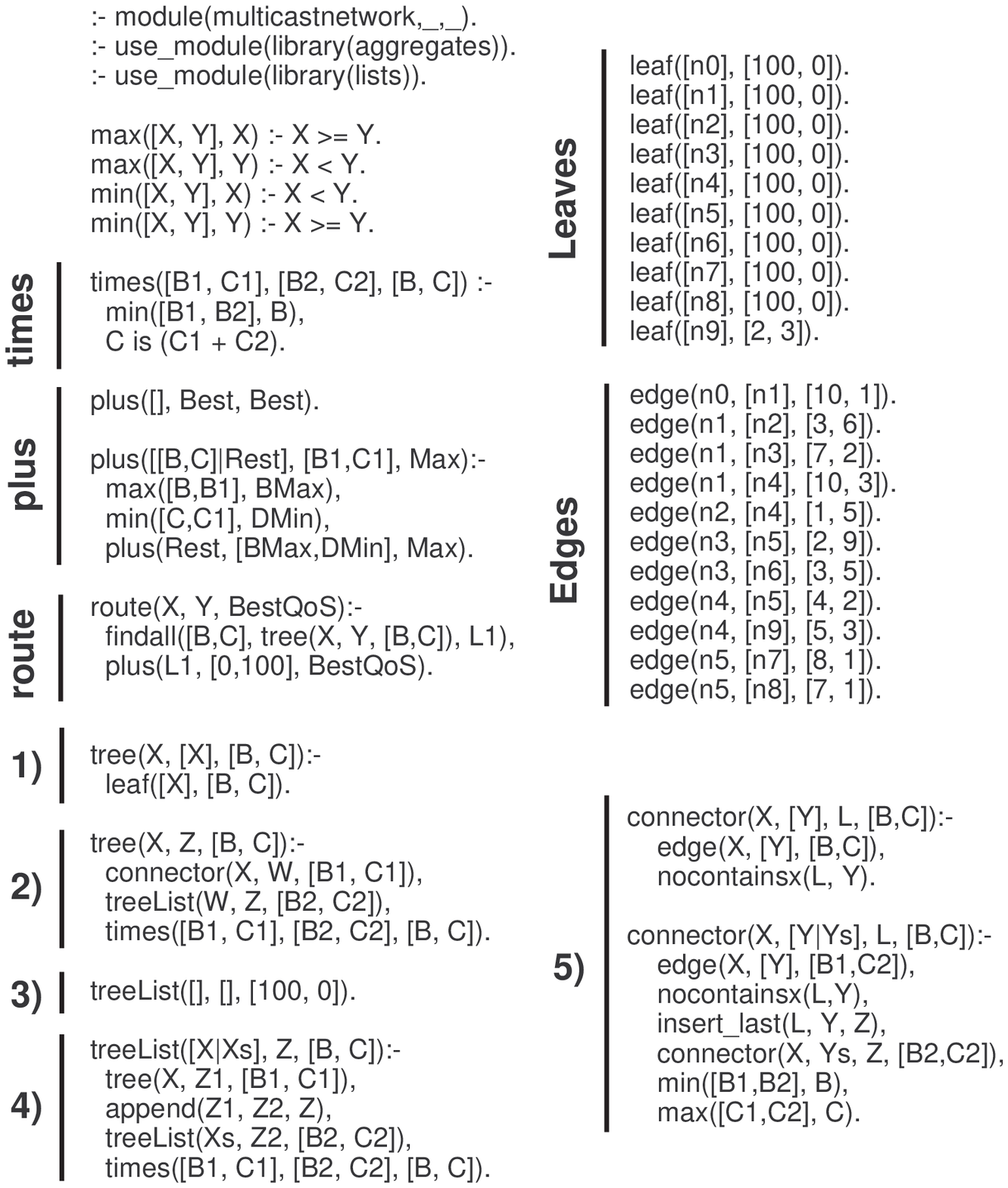}

\end{table}

\begin{figure}
  \centering
    \includegraphics[scale=.35]{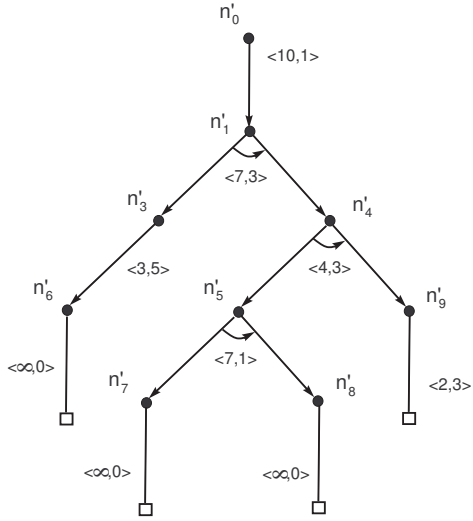}
    \caption{The best multicast distribution tree that can be found with the program in Table~\ref{tab:and-or-graph}.}
    \label{bestMultiTree}
\end{figure}

To solve the \emph{and-or} graph problem it is enough to perform a
query in Prolog language: for example, if we want to compute the
cost of all the trees rooted at $n_0$ and having as leaves the
nodes representing all the receivers (i.e. $\{ n_{6}, n_{7},
n_{8}, n_{9} \}$), we have to perform the query $route(n_{0},
[n_{6},n_{7},n_{8},n_{9}], [B, C])$, where $B$ and $C$ variables
will be instantiated with the bandwidth and cost of the found
trees. The output of the CIAO program for this query corresponds
to the cost of the tree in Fig.~\ref{bestMultiTree}, i.e. $\langle
2,16 \rangle$. For this query, the output of the program in
Tab.~\ref{tab:and-or-graph} is shown in
Fig.~\ref{fig:outputprogram}. The tree in Fig.~\ref{bestMultiTree}
is a \emph{solution tree} (see Sec.~\ref{sec:hypertranslation})
for the graph in Fig.~\ref{translation}\emph{b}, with mapping
function $g: g(n'_0)=n_0, g(n'_1)=n_1, g(n'_3)=n_3, g(n'_4)=n_4,
g(n'_5)=n_5, g(n'_6)=n_6, g(n'_7)=n_7, g(n'_8)=n_8, g(n'_9)=n_9$.

\begin{figure}
\centering
\includegraphics[scale=0.63]{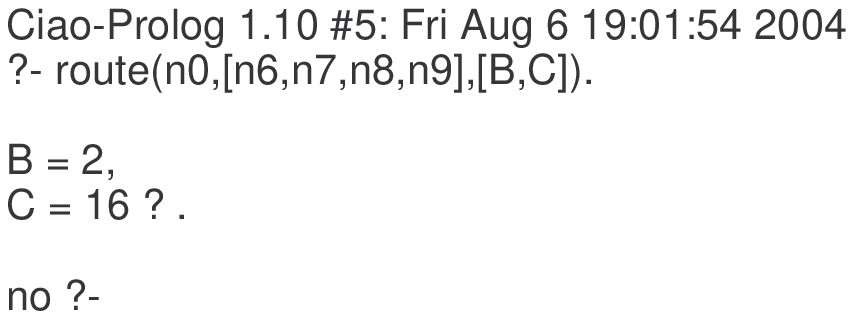}
\caption{The CIAO output for the program in
Tab.~\ref{tab:constrainedpath}. The best bandwidth and delay
values are found for the tree with $n_{6}$,$n_{7}$,$n_{8}$,$n_{9}$
destinations} \label{fig:outputprogram}
\end{figure}

A global cost can be given to \emph{and} trees: recursively, to
every subtree of $H$ with root node $n_{i_0}$, a cost $c_{i_0}$ is
given as follows:
\begin{itemize}
\item If $n_{i_0}$ is a leaf, then its cost is the associated
  constant.
\item If $n_{i_0}$ is the input node of a connector ($n_{i_0}, n_{i_1},
  \ldots, n_{i_k}$), then its cost is $c_{i_0} = f_r(c_{i_1}, \ldots,
  c_{i_k}$) where $f_r$ is the function cost associated with the
  connector, and $c_{i_1}, \ldots, c_{i_k}$ are the costs of the
  subtrees rooted at nodes $n_{i_1}, \ldots, n_{i_k}$.
\end{itemize}

The final cost of the tree in Fig.~\ref{bestMultiTree} obtained
with the CIAO program is equivalent to the one that can be
computed by using $\times'$ to define the $f_r$ cost function.
Starting from the $n'_0$ source node and the connector
$(n'_0,n'_1)$ with cost $\langle 10,1 \rangle$, the total cost of
the tree $c_{n'_0}$ is
$$c_{n'_0} = f_r(c_{n'_1}) =  \langle 10,1 \rangle \times'
c_{n'_1}$$

Clearly, this framework can be used to solve the unicast problem
as well, if the asked query include only one destination node,
e.g. $route(n_{0}, [n_{6}], [B, C])$.

\subsection{Modality-based Steiner Tree Problems}
\label{sec:modality-trees} In this Section, as we provide in
Sec.~\ref{sec:mo} for plain paths, we improve the tree search by
including the possibility of considering some modalities
associated with the use of the hyperarcs. Even in this case the
justification is easy, since sometimes it may be useful to
associate with each hyperarc also the information about the
modality to be used to traverse that specific hyperarc. In this
Section, we show an example using only two of the three modalities
of Sec.~\ref{sec:mo}:  wired link with no encryption service
(\emph{w}), and wireless link with no encryption service
(\emph{l}). Other classes could collect slices of day time in
which network links are preferred to be used (e.g. to better
support the peaks of traffic), or label special conditions of use,
e.g. to support ``night back-up'' or ``black-out'' events.

\begin{figure}
  \centering
    \includegraphics[scale=0.48]{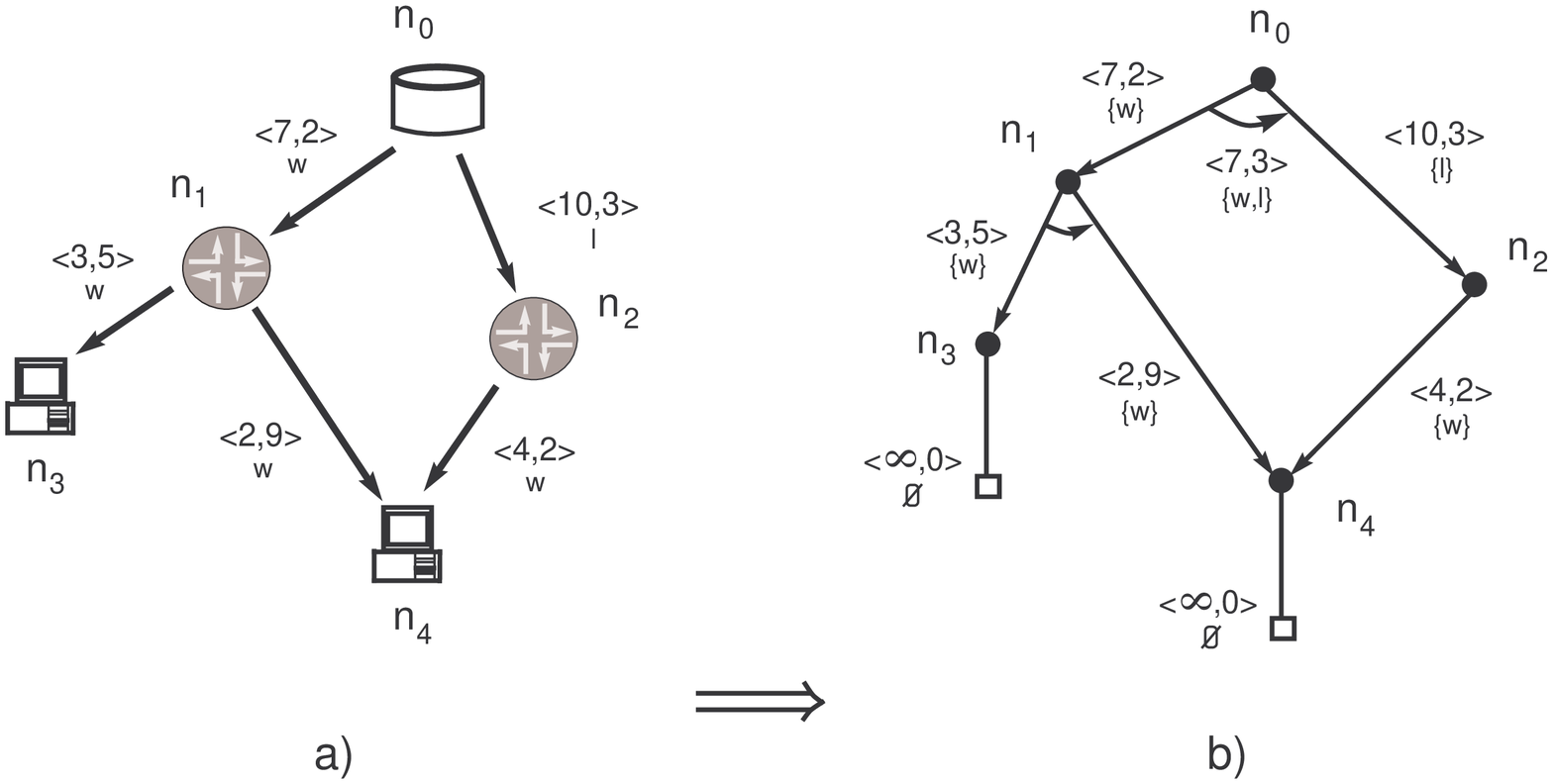}
    \caption{\emph{a)} A network with modalities associated to the links, and \emph{b)} the corresponding hypergraph.}
    \label{fig:multimodality}
\end{figure}

In Fig.~\ref{fig:multimodality} we show an example on how to pass
from a network to a corresponding hypergraph with modalities (from
Fig.~\ref{fig:multimodality}\emph{a} to
Fig.~\ref{fig:multimodality}\emph{b}): the modality associated
with a connector is found by using the union operator (i.e.
$\cup$) on the sets of modalities associated with each of the
links represented by that connector. In the example of
Fig.~\ref{fig:multimodality}, $0$-connectors have emptyset as
label, since in this case we do not need any further information
to finally reach a receiver; however, in general $0$-connector
labels may contain the same modalities as the other $n$-connector
labels, e.g. when they represent the internal structure of a sub
network, as ($n_9$) in Fig.~\ref{translation}\emph{b}. For
example, if the connection from $n_0$ to $n_1$ is wired, and the
connection from $n_0$ to $n_2$ is wireless, the connector
$(n_0,n_1,n_2)$ will be labelled with the $\{w,l\}$ modality set.
Thus, edges are now represented in the following way:

$$edge(source\_node, [dest\_nodes], [bandwidth, cost], [list\_of\_modalities])$$

The query for the tree search must now be performed by including
also the set of allowed modalities: if the set of modalities
associated with a connector is a subset of the modalities asked in
the query, then that connector can be used to build the tree. This
can be practically accomplished by using, for example, the CIAO
\emph{difference} predicate between the two lists (sets) of
modalities, or the \emph{sublist} property.

For example (please refer to Fig.~\ref{fig:multimodality}), asking
for $route(n_{0}, [n_{3},n_{4}], [B, C], [w])$ means that we are
looking for paths made only with wired links (i.e. \emph{w}). The
$(n_0, n_1, n_2)$ connector cannot be used because its label is
$\{w,l\}$ and we do not want to use wireless links (we remind that
\emph{l} stands for wireless link with no encryption service). To
include also that specific connector in the search, we have to ask
the query $route(n_{0}, [n_{3},n_{4}], [B, C], [w,l])$. Clearly,
the final $0$-connectors are always included in trees because they
have an emptyset label.

\section{A Last Refinement on Semirings for Partially-Ordered Problems}
\label{sec:partiallyref}

As seen in Sec.~\ref{sec:po} and~\ref{sec:hypertranslation}, the
costs on the connectors can be represented by vectors of costs,
representing the QoS metric values of the network links. However,
since we can have a partial order, two such pairs may possibly be
incomparable, and this may lead to a strange situation while
computing the semantics of a given goal. Considering the example
in Sec.~\ref{sec:po} and the related program in
Table~\ref{tab:multispp}, if we want to compute the cost and delay
of the best path from $p$ to $v$, by giving the query {\tt :- p.},
the answer in this case is the value $\langle 7,7 \rangle$. While
the semiring value obtained in totally ordered SCLP programs
represents the cost of one of the shortest paths, here it is
possible that there are no routes with this cost: the obtained
semiring value is in fact the greatest lower bound (w.r.t. both
cost and delay) of the costs of all the paths from $p$ to $v$.
This behavior comes from the fact that, if different refutations
for the same goal have different semiring values, the SCLP
framework combines them via the $+$ operator of the semiring
(which, in the case of our example, is the $min'$ operator of
Sec.~\ref{sec:po}). If the semiring is partially ordered, it may
be that $a+b$ is different from both $a$ and $b$. On the contrary,
if we have a total order $a+b$ is always either $a$ or $b$.

This problem of course is not satisfactory, because usually one
does not want to find the greatest lower bound of the costs of all
paths from the given node to the destination node, but rather
prefers to have one of the non-dominated paths. To solve this
problem, we can add variables to the SCLP program, as we did in
the previous Section, and also change the semiring. In fact, we
now need a semiring which allows us to associate with the source
node the set of the costs of all non-dominated path from there to
the destination node. In other words, starting from the semiring
$S = \langle A, +, \times, \0,\1 \rangle$ (which, we recall, in
the example of Sec.~\ref{sec:po} is $\langle \mathbb{N}^2, min',
+', \langle +\infty, +\infty \rangle, \langle 0, 0 \rangle
\rangle$), we now have to work with the semiring $P^H(S) = \langle
P^H(A), \uplus, \times^*, \emptyset,A \rangle$, where:

\begin{itemize}
\item $P^H(A)$ is the Hoare Power Domain~\cite{SMYTH78} of $A$, that is,
$P^H(A) = \{S \subseteq A \mid x \in S, y \leq_S x $ implies $ y
\in S \}$. In words, $P^H(A)$ is the set of all subsets of $A$
which are downward closed under the ordering $\leq_S$. It is easy
to show that such sets are isomorphic to those containing just the
non-dominated values. Thus in the following we will use this more
compact representation for efficiency purposes. In this compact
representation, each element of $P^H(A)$ will represent the costs
of all non-dominated paths from a node to the destination node;

\item the top element of the semiring is the set $A$
(its compact form is $\{\1\}$, which in our example is $\{\langle
0, 0 \rangle \}$);

\item the bottom element is the empty set;

\item the additive operation $\uplus$ is the {\em formal union}~\cite{SMYTH78} that takes two sets and obtains their union;

\item the multiplicative operation $\times^*$ takes two sets and
produces another set obtained by multiplying (using the
multiplicative operation $\times$ of the original semiring, in our
case +') each element of the first set with each element of the
second one;

\item the partial order of this semiring is as follows:
$a \leq_{P^H(S)} b$ iff $a \uplus b = b$, that is for each element
of $a$, there is an element in $b$ which dominates it (in the
partial order $\leq_S$ of the original semiring).
\end{itemize}

From the theoretical results in~\cite{SMYTH78}, adapted to
consider c-semirings, we can prove that $P^H(S)$ and its more
compact form are indeed isomorphic. Moreover, we can also prove
that given a c-semiring $S$, the structure $P^H(S)$ is a
c-semiring as well~\cite{sclpjheu}.

\begin{theorem}\label{theo:hoare}
Given a c-semiring $S = \langle A, +, \times, \0, \1 \rangle$, the
structure $P^H(S)=\langle P^H(A), \uplus, \times^*, \emptyset, A
\rangle$ obtained using the {\em Power domain of Hoare} operator
is a c-semiring.
\end{theorem}
\begin{proof}
The proof easily follows from the properties of the $\times$
operator in the c-semiring $S$ and from the properties
(commutativity, associativity, and idempotence) of the formal
union $\uplus$ in $P^H(S)$.
\end{proof}

Note that in this theorem we do not need any assumption over the
c-semiring $S$. Thus the construction of $P^H(S)$ can be done for
any c-semiring $S$. Notice also that, if $S$ is totally ordered,
the c-semiring  $P^H(S)$ does not give any additional information
w.r.t. $S$. In fact, if we consider as a single element element
the empty set (with the meaning that there are no paths) and the
set containing only the bottom of $A$ (with the meaning that there
exists a path whose cost is $\infty$), it is possible to build an
isomorphism between $S$ and $P^H(S)$ by mapping each element $p$
(a set) of $P^H(A)$ onto the element $a$ of $A$ such that $a \in
p$ and $a$ dominates all elements in the set $p$.

The only change we need to make to the program with variables, in
order to work with this new semiring, is that costs now have to be
represented as singleton sets. For example, clause {\tt $c_{pq}$
:- < 2, 4 >.} will become {\tt $c_{pq}$ :- \{< 2, 4 >\}.}

Still considering the example in Sec.~\ref{sec:po}, Let us now see
what happens in our example if we move to this new semiring. First
we give a goal like {\tt :- p(X)}. As the answer, we get a set of
pairs, representing the costs of all non-dominated paths from $p$
to $v$. All these costs are non-comparable in the partial order,
thus the user is requested to make a choice. However, this choice
could identify a single cost or also a set of them. In this second
case, it means that the user does not want to commit to a single
path from the beginning and rather prefers to maintain some
alternatives. The choice of one cost of a specific non-dominated
path will thus be delayed until later. Other considerations on
this semiring are given in~\cite{sclpjheu}.

Most classical methods to handle multi-criteria SP problems find
the shortest paths by considering each criteria separately, while
our method deals with all criteria at once. This allows to obtain
optimal solutions which are not generated by looking at each
single criteria. In fact, some optimal solutions could be
non-optimal in each of the single criteria, but still are
incomparable in the overall ordering. Thus we offer the user a
greater set of non-comparable optimal solutions. For example, by
using a cost-delay multi-criteria scenario, the optimal solution
w.r.t. cost could be $10${\euro} (with a delay of $100$msec),
while the optimal solution w.r.t. delay could be $10$msec (with a
cost of $100${\euro}). By considering both criteria together, we
could also obtain the solution with $20$ euro and $20$msec!

Note that the considerations on partially-ordered problems of this
Section clearly state for the multicast tree example in
Sec.~\ref{sec:hypertranslation} as well. In this case, $P^H(S) =
\langle P^H(A), \uplus, \times^*, \emptyset,A \rangle$ uses the
semiring for bandwidth-delay multi criteria: $S_{Network} =
\langle \langle {\cal B} \cup \{0, +\infty\}, \mathbb{N} \rangle,
+', \times', \langle 0, +\infty \rangle, \langle +\infty, 0
\rangle \rangle$, where ${\cal B}$ is the set of bandwidth values,
$+'$ is $\langle\mbox{max}, \mbox{min}\rangle$ and $\times'$ is
$\langle\mbox{min}, \mbox{+}\rangle$. Therefore, $\times^*$ uses
$\langle\mbox{min}, \mbox{+}\rangle$ ($\times'$) to compose two
sets, and $\uplus$ the ordering $\leq_S$ defined by
$\langle\mbox{max}, \mbox{min}\rangle$ ($+'$).

Finally, this method is applicable not only to the multi-criteria
case, but to any partial order, giving us a general way to find a
non-dominated path in a partially-ordered SP problem. It is
important to notice here the flexibility of the semiring approach,
which allows us to use the same syntax and computational engine,
but on a different semiring, to compute different objects.

\subsection{Limiting the Number of Partially Ordered Solutions}\label{sec:expsolution}

As presented in Sec.~\ref{sec:partiallyref}, we can use the Hoare
Power Domain operator to retrieve the set of all the non-dominated
paths (unicast) or trees (multicast) when we suppose that our
network links have multiple and incomparable costs (e.g.
bandwidth, cost and delay). This set of solutions is called the
\emph{Pareto frontier} and is guaranteed to contain all optimal
solutions: all the solutions in this set are equivalently
feasible. In other words, the Pareto frontier exactly captures the
available trade-offs between the different QoS objectives.
However, the use of partially ordered structures leads to the
generation of a potentially exponential number of undominated
solutions. When several of different paths/trees exist between the
source and the receiver(s), it is therefore crucial to keep the
number of configurations as low as possible through some form of
approximation. However, the Hoare Power Domain operator can still
be applied in case the sets of QoS costs have few elements and we
really do not know how to refine the search, or if we know that
few routes exist among nodes (these hypotheses limit the number of
solutions).

We do not want to completely deviate from the incomparability
property of the QoS metrics by adopting a total order, otherwise
all the costs could be rephrased as a single one for each link,
making the problem much more easier (e.g. the unicast problem can
be solved in polynomial time~\cite{Cormen}) and less interesting,
as explained in Sec.~\ref{sec:qosrouting}.

The proposed solution consists in avoiding a pointwise comparison
of the single orderings representing the different criteria for
the $+$ operation of the semiring: we instead adopt a function
that composes all the criteria in a single one and then chooses
the best tuple of costs according to a total ordering. Each of the
QoS criteria is composed by using a different importance value
(i.e. a weight $w_i$). Theorem~\ref{theo:weightedHoare} proves
that such function is still a valid $+$ semiring operation for an
\emph{Ordered} Cartesian product of \emph{Weighted}
semirings~\cite{bistabook,jacm97}, i.e. $\langle
\mathbb{R}^{+},min,\hat{+},+\infty,0 \rangle$ (where $\hat{+}$ is
the arithmetic sum):

\begin{theorem}\label{theo:weightedHoare}
Given two \emph{Weighted} semirings $S_1$ and $S_2$ and a relative
preference for their element sets, i.e. $w_1,w_2 \in
\mathbb{R}^{+}$, we define the \emph{Ordered} Cartesian product of
$S_1$ and $S_2 \equiv S_{f} = \langle \langle \mathbb{R}^{+}
\times \mathbb{R}^{+}\rangle, \, $f$ \,, \langle \hat{+} ,\hat{+}
\rangle, \langle +\infty, +\infty\rangle, \langle 0, 0\rangle
\rangle$. Given $\langle a_1, b_1\rangle, \langle a_2, b_2 \rangle
\in \langle \mathbb{R}^{+} \times \mathbb{R}^{+}\rangle$, $f$
(i.e. the $+$ of the semiring) is defined as:
$$ f\,( \langle a_1, b_1\rangle, \langle a_2, b_2\rangle) =
\begin{cases}
\langle a_1, b_1\rangle & \text{\;  if  \;} w_1 a_1 \hat{+} w_2 b_1 > w_1 a_2 \hat{+}  w_2 b_2\\
\langle min(a_1, a_2), min(b_1,b_2) \rangle & \text{\;  if \;} w_1
a_1 \hat{+} w_2
b_1 = w_1 a_2 \hat{+}  w_2 b_2\\
\langle a_2, b_2\rangle & \text{\;  if \;} w_1 a_1 \hat{+}  w_2
b_1 < w_1 a_2 \hat{+}  w_2 b_2\end{cases}
$$
Then $S_f$ is a c-semiring.

\end{theorem}

\begin{proof}
Since the only change w.r.t. a classical Cartesian product of
\emph{Weighted} semirings is the $+$ operator of $S_f$, we only
need to check the properties of $+$ given in
Sec.~\ref{sec:softCLP}. The function $f$ is commutative,
associative, closed, idempotent, $\langle +\infty, +\infty
\rangle$ is its unit element and $\langle 0, 0 \rangle$ its
absorbing element: these properties easily follows from the
properties of $min$ and arithmetic sum and multiplication, which
describe the $f$ expression. We only prove that $\times$ still
distributes over $+$ ($a_i, b_i, w_i \in \mathbb{R}^{+}$):
$$\langle a_1, b_1 \rangle \times (\langle a_2, b_2 \rangle +
\langle a_3, b_3 \rangle) =  \begin{cases} \langle (a_1 \hat{+}
a_2),
(b_1 \hat{+} b_2) \rangle & \text{\;  if  \;} cond_1\\
\langle (a_1 \hat{+} min(a_2,a_3),
(b_1 \hat{+} min(b_2, b_3) \rangle & \text{\;  if  \;} cond_2\\
\langle (a_1 \hat{+} a_3), (b_1 \hat{+} b_3) \rangle & \text{\;
if  \;} cond_3
\end{cases}$$
\vspace{+0.3cm}
$$(\langle a_1, b_1 \rangle \times \langle a_2, b_2 \rangle)
+ (\langle a_1, b_1 \rangle \times \langle a_3, b_3 \rangle) =$$
\vspace{-0.51cm}
$$
\begin{cases} \langle (a_1 \hat{+} a_2),
(b_1 \hat{+} b_2) \rangle &  \text{\;  if  \;} cond_4\\
\langle min(a_1 \hat{+} a_2, a_1 \hat{+} a_3),
min(b_1 \hat{+} b_2, b_1 \hat{+} b_3) \rangle  &  \text{\;  if  \;} cond_5\\
\langle (a_1 \hat{+} a_3), (b_1 \hat{+} b_3) \rangle & \text{\; if
\;} cond_5
\end{cases}$$

Where $cond_1$ is $w_1 a_2 \hat{+} w_2 b_2 > w_1 a_3 \hat{+} w_2
b_3$ and $cond_4$ is $w_1 (a_1 \hat{+} a_2) \hat{+} w_2 (b_1
\hat{+} b_2) > w_1 (a_1 \hat{+} a_3) \hat{+} w_2 (b_1 \hat{+}
b_3)$; by simplifying both sides of $cond_4$ we obtain that
$cond_1 \equiv cond_2$. In the same way we can prove that $cond_2
\equiv cond_5$ if and $cond_3 \equiv cond_6$. Therefore, $\times$
distributes over $+$.\end{proof}

Notice that the proof can be easily extended for an \emph{Ordered}
Cartesian product of $n > 2$ \emph{Weighted} semirings. Notice
also that we can assemble an \emph{Ordered} Cartesian product even
for $n$ \emph{Probabilistic}
semirings~\cite{bistabook,jacm97}, and even for semirings in general, as claimed in Theo.~\ref{theo:weightedHoare2}: 

\begin{theorem}\label{theo:weightedHoare2}
We consider two identical semirings $S_1, S_2 = \langle A, +,
\times, \0, \1 \rangle$ where $\times$ is
cancellative~\cite{ecai06}. We can define an \emph{Ordered}
Cartesian product $S_f$ as $\langle \langle A \times A \rangle, \,
f \,, \langle \times ,\times \rangle, \langle \0, \0 \rangle,
\langle \1, \1 \rangle \rangle$, where $f$ is defined as:
$$ f\,( \langle a_1, b_1\rangle, \langle a_2, b_2\rangle) =
\begin{cases}
\langle a_1, b_1\rangle & \text{\;  if  \;}  a_1 \times  b_1 >_{S_{1,2}}  a_2 \times  b_2\\
\langle a_1 + a_2, b_1 + b_2 \rangle & \text{\;  if \;} a_1 \times
b_1 =_{S_{1,2}} a_2 \times  b_2\\
\langle a_2, b_2\rangle & \text{\;  if \;} a_1 \times b_1
<_{S_{1,2}} a_2 \times b_2\end{cases}
$$

Then $S_f$ is a c-semiring.
\end{theorem}

\begin{proof}
Notice that we use the same $+$ and $\times$ operators of
$S_1,S_2$ also in the definition of $f$, thus their properties
still hold. For this reason, we can easily prove that the $+$ (as
defined by $f$) of the semiring is commutative, associative,
closed, idempotent, $\langle \0, \0 \rangle$ is its unit element
and $\langle \1, \1 \rangle$ its absorbing element. The
cancellative property is needed to prove the that $\times$
distributes over $+$:
$$\langle a_1, b_1 \rangle \times (\langle a_2, b_2 \rangle +
\langle a_3, b_3 \rangle) =  \begin{cases} \langle (a_1 \times
a_2),
(b_1 \times b_2) \rangle & \text{\;  if  \;} cond_1\\
\langle (a_1 \times (a_2 + a_3), (b_1 \times (b_2+ b_3)) \rangle &
\text{\; if
\;} cond_2\\
\langle (a_1 \times a_3), (b_1 \times b_3) \rangle & \text{\; if
\;} cond_3
\end{cases}$$ \vspace{+0.3cm} $$(\langle a_1, b_1 \rangle \times \langle a_2, b_2 \rangle)
+ (\langle a_1, b_1 \rangle \times \langle a_3, b_3 \rangle)
=$$\vspace{-0.51cm}
$$
\begin{cases} \langle (a_1 \times a_2),
(b_1 \times b_2) \rangle & \text{\;  if  \;} cond_4\\
\langle (a_1 \times a_2) + (a_1 \times a_3),
(b_1 \times b_2) + (b_1 \times b_3) \rangle & \text{\;  if  \;} cond_5\\
\langle (a_1 \times a_3), (b_1 \times b_3) \rangle & \text{\;  if
\;} cond_6
\end{cases}$$

Where $cond_1$ is $a_2 \times b_2 >_{S_{1,2}} a_3 \times b_3$ and
$cond_4$ is $(a_1 \times a_2)  \times (b_1 \times b_2)
>_{S_{1,2}}  (a_1 \times a_3)  \times (b_1 \times b_3)$. Since
$\times$ is cancellative, we can simplify both sides of $cond_4$
and we obtain that $cond_1 \equiv cond_4$. In the same way we can
prove that $cond_2 \equiv cond_5$ if and $cond_3 \equiv cond_6$.
Therefore, $\times$ distributes over $+$.

\end{proof}

With Theo.~\ref{theo:weightedHoare} and
Theo.~\ref{theo:weightedHoare2} we show that multiple semirings of
the same type (e.g. \emph{Weighted} or \emph{Probabilistic}) can
be composed together according to some expressed preferences. In
this way, the resulting tuples are totally ordered and the final
solution consists in the most preferred one. Ad-hoc compositions
can be used also to merge different semirings in a single one,
e.g. \emph{Weighted} and \emph{Probabilistic}. However, according
to the definition of $f$ in Theo.~\ref{theo:weightedHoare2}
(similar considerations hold for Theo.~\ref{theo:weightedHoare}),
$f\,( \langle a_1, b_1\rangle, \langle a_2, b_2\rangle) = \langle
a_1 + a_2, b_1 + b_2\rangle$ if $a_1 \times b_1 =_{S_{1,2}} a_2
\times  b_2$, and thus $f$ returns the lowest upper bound of the
two couples. As already said in Sec.~\ref{sec:partiallyref}, this
result does not represent a ``real'' solution. Nonetheless, this
problem can be overcome by collecting all the best equivalent
couples in the same set, i.e. applying the Hoare Power Domain
operator (see Sec.~\ref{sec:partiallyref}).

\begin{corollary}\label{theo:weightedHoare3}Given an \emph{Ordered}
Cartesian product $S_f = \langle \langle A \times A \rangle, \, f
\,, \langle \times ,\times \rangle,$ $\langle \0, \0 \rangle,
\langle \1, \1 \rangle \rangle$ as described in
Theo.~\ref{theo:weightedHoare} and the Hoare Power Domain operator
$P^H$, then $P^H(S_f)$ is a semiring.
\end{corollary}
\begin{proof}
Given the results in Theo.~\ref{theo:hoare} (see
Sec.~\ref{sec:partiallyref}), we can easily assemble the Hoare
Power Domain over the semiring proposed in
Theo.~\ref{theo:weightedHoare2}, by using the Hoare Power Domain
operator (see Sec.~\ref{sec:partiallyref}).
\end{proof}

A similar result can be proved for the semiring assembled with
Theo.~\ref{theo:weightedHoare} (i.e. for the \emph{Weighted}
semirings), by applying to it the Hoare Power Domain operator as
well.

\section{Solving the problem in practice}\label{sec:solving}

\subsection{Scale-free Networks}\label{app:small} Small-world
networks may belong to three classes: single-scale, broad-scale,
or scale-free depending on their connectivity distribution $P(k)$,
which is the probability that a randomly selected node has exactly
$k$ edges. Scale-free networks follow a power law of the generic
form $P(k) \backsim k^{-\gamma}$~\cite{powerlaw}: in words, in
these networks some nodes act as ``highly connected hubs'' (with a
high degree), although most nodes are of low degree. Intuitively,
the nodes that already have many links are more likely to acquire
even more links when new nodes join in the graph: this is the
so-called ``rich gets richer'' phenomenon. These hubs are the
responsible for the small world phenomenon. The consequences of
this behavior are that, compared to a random graph with the same
size and the same average degree, the average path length of the
scale-free model is somewhat smaller, and  the clustering
coefficient of the network is higher, suggesting that the graph is
partitioned in sub-communities.

Several works as~\cite{powerlaw,as-ir} show that Internet topology
can be modeled with scale-free graphs: in~\cite{as-ir} the authors
distinguish between the \emph{Autonomous System} (AS) level, where
each AS refers to one single administrative domain of the
Internet, and the \emph{Internet Router} level (IR). At the IR
level, we have graphs with nodes representing the routers and
links representing the physical connections among them; at the AS
level graphs each node represents an AS and each link represents a
peer connection trough the use of the \emph{Border Gateway
Protocol} (BGP) protocol. Each AS groups a generally large number
of routers, and therefore the AS maps are in some sense a
coarse-grained view of the IR maps. The same authors
of~\cite{as-ir} confirm the scale-free property for both these
kinds of graphs with a $\gamma = 2.1 \pm 0.1$, even if IR graphs
have a power-law behavior smoothed by an exponential cut-off: for
large $k$ the connectivity distribution follows a faster decay,
i.e. we have much less nodes with a high degree. This truncation
is probably due to the limited number of physical router
interfaces. In~\cite{ultrasmall} the authors prove that scale free
networks with $2 < \gamma < 3$ have a very small diameter, i.e.
$\ln \ln N$, where $N$ is the number of nodes in the graph.

Therefore, we decided to test our QoS routing framework on this
kind of networks because they properly model both the AS and the
IR levels.

\subsection{Implementing the Framework}\label{sec:implementation}
\begin{figure}[h]

     \centering\includegraphics[scale=0.35]{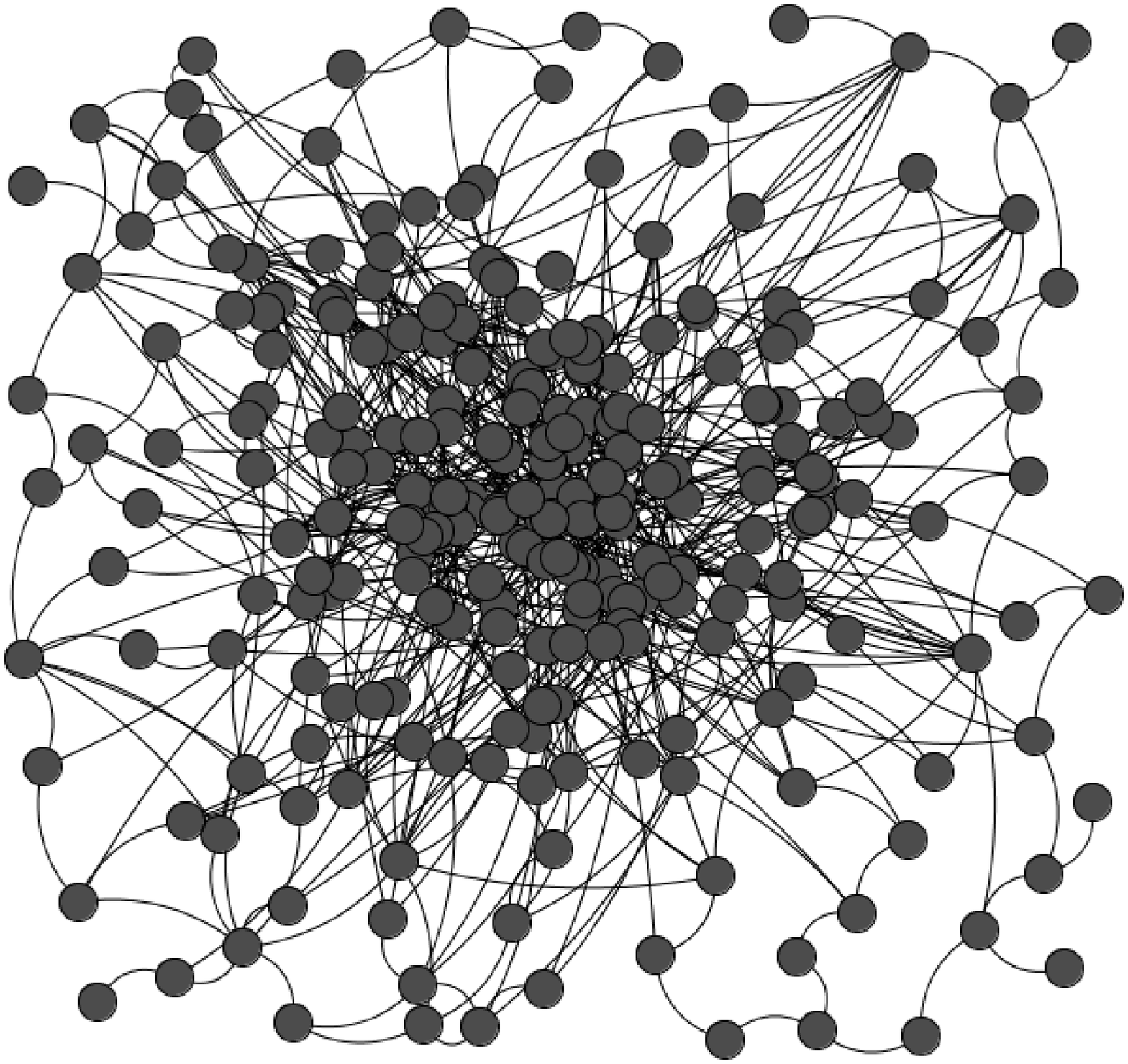}

   \centering\begin{tabular}{| c | c | c | c | c | c | c | c |}
    \hline
    Nodes & Edges & Clustering & Avg. SP & Min Deg & Max Deg. & Avg. Deg & Diameter\\ \hline
    265 & 600 & 0.13 & 3.74 & 1 & 20 & 4.52 & 8 \\ \hline
  \end{tabular}
  \caption{The test scale-free network and the related statistics.}
    \label{fig:as}
\end{figure}

To develop and test a practical implementation of our model, we
adopt the \emph{Java Universal Network/Graph Framework}
(JUNG)~\cite{JUNG}, a software library for the modeling, analysis,
and visualization of  a graph or network. With this library it is
also possible to generate scale-free networks according to the
preferential attachment proposed in~\cite{barabasi}: each time a
new vertex $v_n$ is added to the network $G$, the probability $p$
of creating an edge between an existing vertex $v$ and $v_n$ is $p
= (degree(v) + 1) / (|E| + |V|)$, where $|E|$ and $|V|$ are
respectively the current number of edges and vertices in $G$.
Therefore, vertices with higher degree have a higher probability
of being selected for attachment. We generated the scale-free
network in Fig.~\ref{fig:as} (the edges are undirected) and then
we automatically produced the corresponding program in CIAO (where
the edges are directed), as shown in Sec.~\ref{sec:SCLPandGraphs}.
The reported statistics suggest the scale-free nature of our
network: a quite high clustering coefficient, a low average
shortest path and a high variability of vertex degrees (between
average and max). These features are evidences of the presence of
few big hubs that can be used to shortly reach the destinations.

However, with the CIAO program representing the network in Fig.~\ref{fig:as},
all the queries we tried to perform over that graph were explicitly stopped after $5$ minutes without
discovering the best QoS route solution.
Therefore, a practical implementation definitely needs a strong performance
improvement: in Sec.~\ref{sec:complexity} we show some possible solutions that could all be used also together.
In Sec.~\ref{sec:eclipse} we show an implementation of the exactly same program in ECLiPSe~\cite{eclipse}:
in addition, we use branch-and-bound to prune the search and we claim that only this technique is sufficient to experience a feasible response time
for the queries.

\section{Performance and feasible encodings} \label{sec:complexity}
Although the framework we present in this paper is conceived as a
declarative and expressive mean to represent the QoS routing
problem, some of the used encodings represent an obstacle towards
a real use on practical  cases. Our study is clearly not
aimed at a successful performance comparison with dedicated
algorithms running inside routers or network devices: we instead
desire to model many different routing constraints (e.g. routing and policy constraints)
all inside the same framework. The power of this model is in the facility
with which routing constraints and network bounds in general can be expressed
and added to pre-existing rules. However, we need also a feasible implementation to
obtain and check a solution for real-case networks, even if not
performing as well as the algorithms reported in Sec.~\ref{sec:unicastbg} and Sec.~\ref{sec:multicastbg}.
All these works are focused only on some metrics (e.g. DVMA~\cite{DVMA} considers only delay and jitter)
or adopts ad-hoc heuristics to relax the problem.
Since in Sec.~\ref{sec:implementation} we
prove that a straightforward implementation is not feasible in
practice, in this Section we provide the methods to lighten these
encodings and tackle down the performance problems. For the reason
that we use a general and open framework, we will suggest general strategies to enhance the results.
However, we think that more specific techniques can be used as well.

Notice that the techniques we are going to present but not directly implement in practice
(as \emph{tabling} in Sec.~\ref{sec:complexitysub}) for sake of brevity, have however a strong and accepted
background concerning their efficiency.

\subsection{Using a cut function to reduce the number of solutions}\label{sec:cut}

In Sec.~\ref{sec:expsolution} we solve the (potentially) exponential
space problem linked to the Pareto optimal frontier of the multicriteria
solutions: in that case, we ``flat'' the partial ordering by composing all the criteria together
and using a total ordering on the result.

However, reducing all the QoS costs to a single one is a
coarse simplification that can be applied only in some cases: it is
not always possible to completely rank all the preferences among
themselves (e.g. the ``user'' could not have clear ideas, or it could
not be possible to ``mix'' different metrics together), and it
could be often pleasing to show more results to the final user.
Moreover, as reported in literature, with a single metric the
problem becomes much less interesting: for example, the unicast
problem becomes solvable in polynomial time, instead of to be
NP-Complete (see Sec.~\ref{sec:npcomplete}).

For this reason, in Def.~\ref{def:cutPareto} we define a cut function
that can be applied each time on the result of the formal union
(i.e. $\uplus$) of the Hoare Power Domain defined in
Sec.~\ref{sec:partiallyref}. After the cut, the set contains only
the best tuples of costs, chosen with the criteria defined by the
function. Definition~\ref{def:cutPareto} is presented for the
\emph{Weighted} semirings, but other ad-hoc cuts can be defined for
other types of semirings, just in case the criteria wanted to
reduce the number of solutions cannot be represented with a
semiring-based structure (as in Theo.~\ref{theo:weightedHoare}).
In words, the costs of a tuple $t$ are composed in a single cost
$c_t$ with the aid of a weight for each tuple element: this weight
can change in a predefined interval, thus different $c_t$ can be
obtained. Then, $t$ is deleted from the set if, for each of its
$c_t$, there always exists another tuple $v$ in the set and a cost
$c_v > c_t$.

\begin{definition}\label{def:cutPareto}
We consider a set $P$ of partially ordered $n$-tuples $\langle
a_1, a_2, \dots, a_n \rangle$, where $a_i \in \mathbb{R}^{+}$ in
$\langle \mathbb{R}^{+},min,\hat{+},+\infty,0 \rangle$ (i.e. a
\emph{Weighted} semiring); each $a_i$ is associated  with a weight
$w_i$ in the interval $[k_i - \epsilon_i, k_i + \epsilon_i]$ and
$k_i,\epsilon_i \in \mathbb{R}^{+}$. A cut function can be defined
as $cut(P) = P_{cut} \subseteq P$, where $P_{cut} = \{\langle b_1,
b_2, \dots, b_n \rangle \in P| \nexists \langle c_1, c_2, \dots,
c_n \rangle \in P.(w_1 b_1 \hat{+} w_2 b_2 \hat{+} \dots \hat{+}
w_n b_n) < (w_1 c_1 \hat{+} w_2 c_2 \hat{+} \dots \hat{+} w_n
c_n), \forall w_i \in [k_i - \epsilon_i, k_i], i \in  \{1..n\} \}$
\end{definition}

Therefore, we reduce the number of solutions and we continue considering a partial order and not a total one
(which is important for us, as explained before), but we discard
``bad'' tuples of cost, where ``bad'' is according to the expressed preferences.
Notice  that not all the different criteria must have an associated weight,
and the cut can be performed only considering a subset of metrics.

Notice also that this cut function can be easily modelled with CIAO
Prolog clauses by considering the solutions as lists and by using
the \emph{delete} predicate on the elements that do not satisfy
the given conditions. Notice also that the preference criteria are
different from the ones described in
Theo.~\ref{theo:weightedHoare}: i.e. it can be proved that the
final set of solution obtained with the Hoare Power Domain
operator (see Theo.~\ref{theo:weightedHoare3}) is a subset of the
set found with the cut function in Def.~\ref{def:cutPareto}.

\subsection{Tabled Soft Constraint Logic Programming and Network Decomposition} \label{sec:complexitysub}
In logic programming, the basic idea behind \emph{tabling} (or
\emph{memoing}) is that the calls to tabled predicates are stored
in a searchable structure together with their proven instances:
subsequent identical calls can use the stored answers without
repeating the computation. This collection of tabled subgoals
paired with their answers, generally referred to as {\em call
table} and {\em answer table} respectively, is consulted whenever
a new call, \emph{C}, to a tabled predicate is issued. If \emph{C
}is  similar to a tabled subgoal \emph{S}, then the set of
answers, \emph{A}, associated with \emph{S} may be used to satisfy
\emph{C}. In such instances, \emph{C} is resolved against the
answers in \emph{A}, and hence we refer to \emph{C} as a
\emph{consumer} of \emph{A} (or \emph{S}). If there is no such
\emph{S}, then \emph{C} is entered into the call table and is
resolved against program clauses. As each answer is derived during
this process, it is inserted into the answer table entry
associated with \emph{C} if it contains information not already in
\emph{A}. Furthermore, left recursion need not lead to
non-termination because identical subgoals are not evaluated, and
thus the possible infinite loops are avoided.

Tabling improves the computability power of Prolog systems and for
this reason many programming frameworks have been extended in this
direction. Due to the power of this extension, many efforts have
been made to include it also in CLP, thus leading to the
\emph{Tabled Constraint Logic Programming} (TCLP) framework.
In~\cite{Tab2000} the authors present a TCLP framework for
constraint solvers written using attributed variables; however,
when programming with attributed variables, the user have to take
care of of many implementation issues such as constraint store
representation and scheduling strategies. A more recent
work~\cite{Tab2004} explains how to port \emph{Constraint Handling
Rules} (CHR) to XSB (acronym of \emph{eXtended Stony Brook}), and
in particular its focus is on technical issues related to the
integration of CHR with tabled resolution: as a result, a CHR library is
presently combined with tabling techniques within the XSB system. CHR is a
high-level natural formalism to specify constraint solvers and
propagation algorithms. This a further promising framework where to solve
QoS routing problems and improve the performance (for example,
tabling efficiency is shown in~\cite{tabperf}), since soft
constraints have already been successfully ported to the CHR
system~\cite{SoftCHR}. Hence, part of the soft constraint solving
can be performed once and reused many times.

\begin{figure}
  \centering
    \includegraphics[scale=0.45]{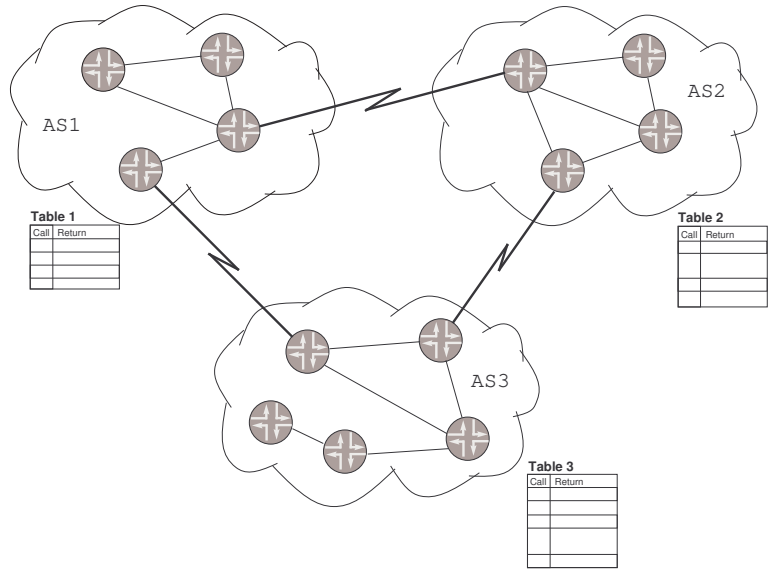}
    \caption{A network subdivided in Autonomous Systems; each AS can store in its border routers a table
    with the goals related to that specific AS.}
    \label{fig:as2}
\end{figure}

One more consideration that can be taken into account while trying
to reduce the complexity, is that large networks, as Internet, are
already partitioned into different \emph{Autonomous System}
(AS)~\cite{MOSPF}, or however, into subnetworks. An AS is a
collection of networks and routers under the control of one entity
(or sometimes more) that presents a common routing policy to the
Internet. AS can be classified by observing the types of traffic
traversing them. A \emph{multihomed} AS  maintains connections to
more than one other AS; however, it would not allow traffic from
one AS to pass through on its way to another AS. A \emph{stub} AS
is only connected to a single AS. A \emph{transit} AS provides
connections through itself to the networks connected to it.
Considering Fig.~\ref{fig:as2}, network \emph{AS1} can use the
transit \emph{AS3} to connect to network \emph{AS2}. An \emph{AS
number} (or ASN) uniquely identifies each AS on the internet (i.e.
\emph{AS1}, \emph{AS2} and \emph{AS3}).

As shown in Fig.~\ref{fig:as2}, in each AS (or subnetwork in
general) we can find a table with the QoS routing goals concerning
the destinations (routers and hosts) within its bounds, by using
tabling techniques. At this point, these tables helps to find the
routes that span multiple ASs and the search procedure is
considerably speeded up: the routes internal to each AS can be
composed together by simply using the links connecting the border
routers. For example, consider when a sender in \emph{AS1} needs
to start a multicast communication towards some receivers in
\emph{AS2} and \emph{AS3}: the routers inside \emph{AS1} can use
\emph{Table 1} to find the routes from the source to the border
routers of \emph{AS1} (i.e. it can communicate with other ASs).
Then, the border routers in \emph{AS2} and \emph{AS3} respectively
use \emph{Table 2} and \emph{Table 3} to find the second and final
part of the route towards the receivers inside their AS. The
procedure of finding such a goal table for a single AS is much
less time consuming than finding it for the whole not-partitioned
network. Clearly, the fundamental premise to obtain a substantial
benefit from this technique is to have strongly-connected
subnetworks and few ``bridges'' among them.

\subsection{An implementation in ECLiPSe} \label{sec:eclipse}

As shown in Sec.~\ref{sec:hypertranslation}, the representation of
the f-star of node in the multicast model can be composed by a
total of $O(2^n)$ connectors, thus in the worst case it is
exponential in the number of graph nodes. This drawback, which is
vigorously perceived in strongly connected networks, and together
with considering a real case network linking hundreds of nodes,
would heavily impact on the time-response performance during a
practical application of our model. Therefore, it is necessary to
elaborate some improvements to reduce the complexity of the tree
search, for example by visiting as few branches of the SCLP tree
as possible (thus, restricting the solution space to be explored).
For this reason, we provide a further implementation by using the
ECLiPSe~\cite{eclipse} system.

ECLiPSe is a software system for the development and deployment of
constraint programming applications, e.g. in the areas of
planning, scheduling, resource allocation, timetabling, transport
and more. It contains several constraint solver libraries, a
high-level modelling and control language, interfaces to
third-party solvers, an integrated development environment and
interfaces for embedding into host environments~\cite{eclipse}. We
decided to use ECLiPSe because of its extendibility and efficiency
due to its wide range of optimization libraries (e.g. on
\emph{symmetry breaking}). In particular, we exploit the
\emph{branch\_and\_bound} library in order to reduce the space of
explored solutions and consequently improve the performance.
Branch-and-bound is a well-known technique for optimization
problems, which is used to immediately cut away not promising
partial solutions, by basing on a ``cost'' function.Unfortunately, as far
as we know, ECLiPSe does not support tabling techniques (introduced in~\ref{sec:complexitysub})
and therefore it cannot be adopted to compose the benefits of both techniques.

\begin{figure}
  \centering
    \includegraphics[scale=0.63]{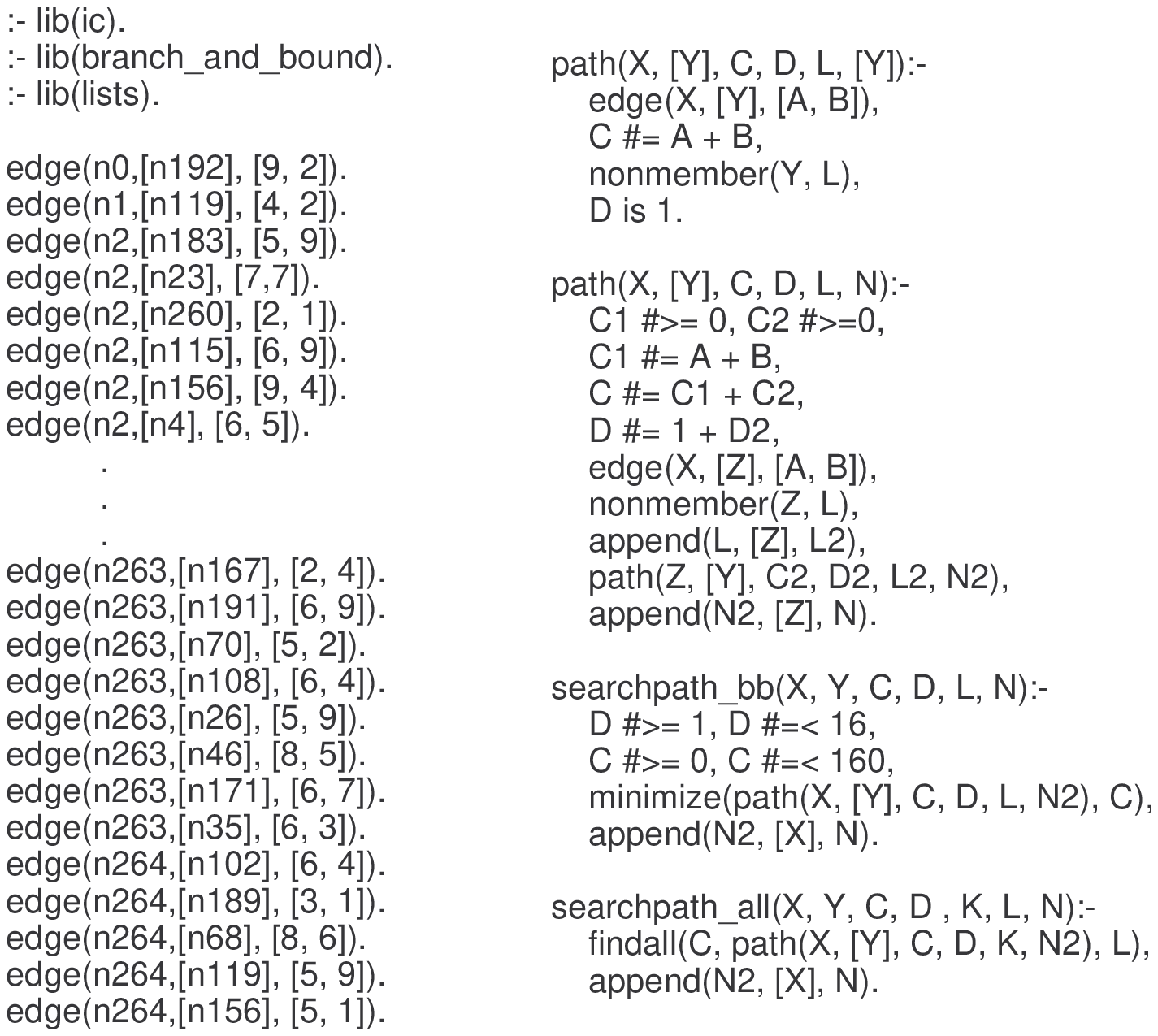}
    \caption{The representation in ECLiPSe (with branch-and-bound optimization) of the QoS routing problem for the network in Fig.~\ref{fig:as};
    clearly, only some of the $600$ edges are shown.}
    \label{fig:eclipse}
\end{figure}

In Fig.~\ref{fig:eclipse} we show a program in ECLiPSe that represents
the unicast QoS routing problem for the scale-free network in Fig.~\ref{fig:as}.
We decided to show only the unicast case for sakes of clarity, but feasible time responses
can be similarly obtained for the multicast case (i.e. searching for a tree instead of a plain path) by
working on the branch-and-bound interval of explored costs, as we will better explain in the following.
Clearly, in Fig.~\ref{fig:eclipse} we report only some of the $600$ edges of the network.

The code in Fig.~\ref{fig:eclipse} has been automatically generated
with a \emph{Java} program using JUNG, as done for the CIAO program in Sec.~\ref{sec:implementation}: the corresponding text
file is $30$Kbyte. The size can be halved by not printing the reverse links and generating them with a specific clause, if each link
and its reverse one have the same cost.

\begin{figure}
  \centering
    \includegraphics[scale=0.45]{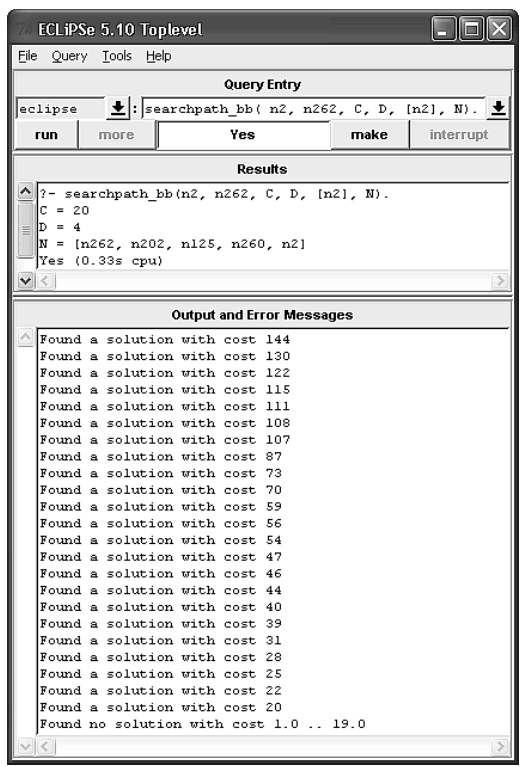}
    \caption{The ECLiPSe shell with the query $searchpath\_bb(n6, n261, C , D, [n6], L)$ and the corresponding found result for the program in Fig.~\ref{fig:eclipse}.}
    \label{fig:eclipse_shell}
\end{figure}

The branch-and-bound optimization is achieved with $minimize(+Goal, ?Cost)$ (importing the
\emph{branch\_and\_bound}  library) in the $searchpath\_bb$ clause in Fig.~\ref{fig:eclipse},
where the \emph{Goal} is a nondeterministic search routine (the clauses that describe the \emph{path}
structure) that instantiates a \emph{Cost} variable (i.e. the QoS cost of the path) when a solution is found.
Notice that for each of the edges of the network we randomly generated two different QoS costs by using the \emph{java.util.Random Class}, each of them in the interval $[1..10]$.
Therefore the cost of link is represented by a couple of values. In order to model the semiring we propose in Theo.~\ref{theo:weightedHoare}, the cost of the path
is computed by summing the two QoS features together (i.e. \emph{A} and \emph{B} in Fig.~\ref{fig:eclipse}):
we compute $w_1 A + w_2 B$ and we suppose $w_1 = w_2 = 1$, i.e. the composed cost of a link is in the interval $[2..20]$.
ECLiPSe natively allows to apply a branch-and-bound procedure focused only on a single cost variable, but ad-hoc techniques can be developed to consider
also the cut function presented in Def.~\ref{def:cutPareto}, in order to keep a real multicriteria preference for the QoS features.

The two clauses $searchpath\_bb$ and $searchpath\_all$ represent the queries that
can be asked to the system: they respectively use and not use the branch-and-bound optimization,
i.e. $searchpath\_all$ finds all the possible paths in order to find the best one.
In order to describe the structure of a $searchpath\_bb$ query (see Fig.~\ref{fig:eclipse}), we take as example $searchpath\_bb(n2, n262, C , D, [n2], L)$: with this query we want to find the best path between the nodes
$n2$ and $n262$, $C$ is the cost of the path (used also by the branch-and-bound pruning), $D$ is the number of hops, $L$  (in Fig.~\ref{fig:eclipse}) is the list of already traversed nodes
and $N$ is a list used to collect the nodes of the path (in reverse order).
The result of this query is reported in Fig.~\ref{fig:eclipse_shell},
by showing directly the ECLiPSe window: the best cost value (i.e. $20$) was
found after $0.33$ seconds with a path of $4$ hops, i.e. $n2$-$n260$-$n125$-$n202$-$n262$.

The corresponding query $searchpath\_all(n6, n261, C , D, K, [n6], N)$ ($K$ is the list of
solutions found by the \emph{findall} predicate), which does not use the branch-and-bound pruning (and constraints),
was explicitly interrupted after $10$ minutes without finding the goal. Other queries are
satisfied in even less than one second, depending on the efficiency of the pruning efficiency for
the specific case.

To better describe and accelerate the search we added also some constraints,
which are explained in Tab.~\ref{tab:constraints}. In Fig.~\ref{fig:eclipse}
we also import the hybrid integer/real interval arithmetic constraint solver
of ECLiPSe to use them, i.e. the \emph{ic} library. Notice that the
constraints depending on the \emph{Diameter} of the network (i.e. $8$, as shown in Fig.~\ref{fig:as}) limit the search space and provides a mild
approximation at the same time: in scale-free networks, the
average distance between two nodes can be $\ln \ln N$, where $N$
is the number of nodes~\cite{ultrasmall} (see also Sec.~\ref{app:small}). Therefore,
considering a max depth of the path as twice the diameter value (i.e. $16$) still
results in a large number of alternative routes, since, for the scale-free
network in Fig.~\ref{fig:as}, this value is $4$-$5$ times the average shortest path of the network (i.e. $3.74$ as shown in Fig.~\ref{fig:as}).

\begin{table}\centering \includegraphics[scale=0.64]{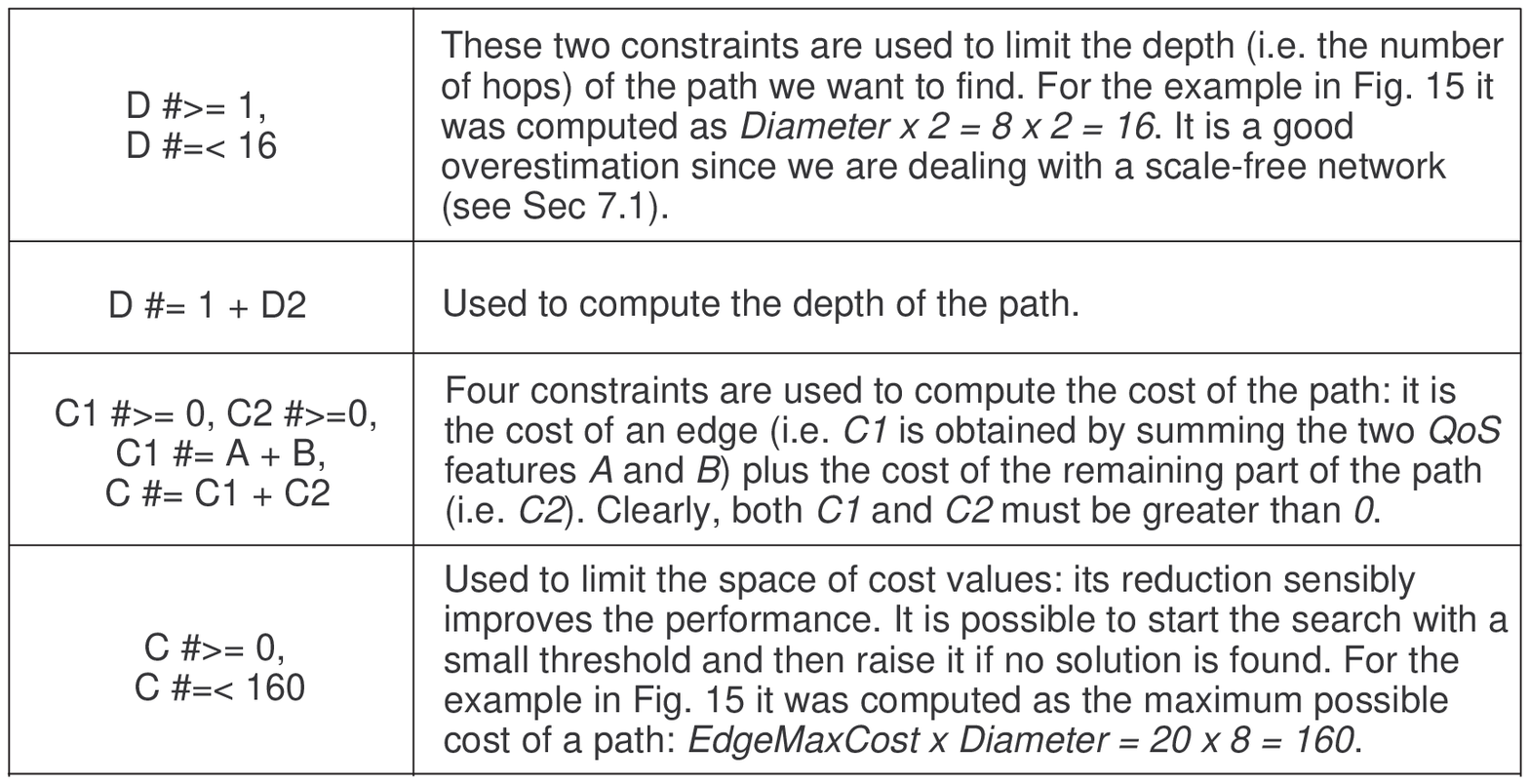}
\caption{The description of the constraints used in Fig.~\ref{fig:eclipse}.}
\end{table}\label{tab:constraints}

In order to show the scalability property of our framework, in Tab.~\ref{table:performance} we summarize the performance results of $50$ queries executed
on three distinct scale-free networks with a different number of nodes: $n=50$, $n=265$ (i.e. the network in Fig.~\ref{fig:as}) and $n=877$.
These statistics are related to the \emph{Min/Max/Average Time} needed to obtain a path, its \emph{Average Cost} and its \emph{Max/Average Depth}.
For each query, the source and destination nodes have been randomly generated. We can see that \emph{Max Time} sensibly differs from the \emph{Average Time}, and this is due to the
poor efficiency of the branch-and-bound pruning in some cases. However, this technique performs very well in most of cases,
as the low \emph{Average Time} Tab.~\ref{table:performance} shows (even for $n=877$).

\begin{table}[h]

   \centering\begin{tabular}{| c | c | c | c | c | c | c |}
    \hline
    Nodes & Min Time & Max Time & Avg. Time & Avg. Cost & Avg. Depth & Max Depth\\ \hline
    50 & \small{$\thicksim$}~0s & 0.45s & 0.1s & 17.54 & 3.04 & 7\\ \hline
    265 & 0.02s & 77.12s & 4.08s & 29.8 & 5.46 & 11\\ \hline
    877 & 0.5s & 40.05s & 4.89s & 37.72 & 6.72 & 14\\ \hline
  \end{tabular}
  \caption{Some performance statistics obtained with the ECLiPSe framework (with branch-and-bound),
  collected on three different size networks (i.e. $50$, $265$ and $1000$ nodes). On each network we performed $50$ queries.}
    \label{table:performance}
\end{table}

Comparable performance results are achievable as well also for the multicast case,
by enforcing the structure of the tree with other ad-hoc constraints: for example, by constraining
the width of the searched tree to the number of the multicast receivers in the query, since it is useless to find wider trees. Moreover, the problem
can be first over-constrained and then relaxed step-by-step if no solution is found. For example, we can start
by searching a solution in the cost interval $[0..35]$ and then, if the best solution
is not included in this interval, setting the interval to $[36..70]$ (and so on until the best solution is found).
Notice that in this way we strongly speed-up the search while preserving all the information, due to the characteristics of
the branch-and-bound technique. This behaviour can be easily reproduced in ECLiPSe, since the customizable
options of $bb\_min(+Goal, ?Cost, ?Options)$ (i.e. another clause to express branch-and-bound) include the $[From..To]$ interval parameters.

At last, we are confident that the ECLiPSe system can be used to further improve the performance,
since it is possible to change the parameters of branch-and-bound, e.g. by changing the strategy after finding a solution~\cite{eclipse}:
\emph{continue} search with the newly found bound imposed on Cost, \emph{restart} or perform a \emph{dichotomic}
after finding a solution, by splitting the remaining cost range and restart search to find a solution in the lower sub-range.
If it fails, the procedure assumes the upper sub-range as the remaining cost range and splits again. Moreover, it is possible to
add \emph{Local Search} to the tree search, and to program specific heuristics~\cite{eclipse}.

\subsubsection{Further reducing the dimension of $n$-connectors} \label{sec:redtopology}

\begin{figure}
  \centering
    \includegraphics[scale=0.6]{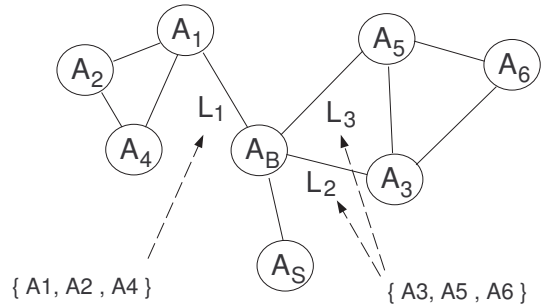}
    \caption{Routing information can be added to the link clauses to
    avoid parts of the network (e.g. the $L_1$ link can be avoided if the destination is $A_6$).}
    \label{fig:routeinfo}
\end{figure}

One more enhancement that can be accomplished to reduce the size of a node's neighborhood
(w.r.t. the given query) for the multicast distribution, is the inclusions of program facts that describe the topology of
the network (or part of it). In this way, like in classic network
routing, we can immediately remove from the search the not
involved clusters or those clusters we do not want to cross for policy reasons. For
example, we can add the list of reachable ASs directly in each connector: if a connector
allows us to reach $\{AS_1, AS_2, AS_4\}$ but not $\{AS_3, AS_5\}$, we can use the first
list as the additional routing information linked to that connector ($A_i$ represent
constant names). If the intersection between the ASs related to the receivers in the
query and the list of a given connector is empty, then we can avoid considering that
connector in the search since it will reach only not interesting nodes.
An graphical example of this behaviour is given in Fig.~\ref{fig:routeinfo}.

Clearly, other hierarchical
partitions can be adopted instead of large ASs: for example we can consider
simple subnetworks if we have to deal with a small departmental networks.
Considering scale-free networks (see Sec.~\ref{app:small}), these improvements are strongly needed
for hub nodes, i.e. the backbone nodes of the network with a high degree: these nodes connect
a lot of separate networks together and thus we can avoid to explore those branched not touched by the query.
This kind of relaxation can be easily programmed in CIAO Prolog, with lists of terms (to represent
the list of ASs reached by a $1$-connector), \emph{union} predicate (to join the lists of $1$-connectors) and
\emph{difference} predicate to check that the AS lists of the query and the obtained $n$-connector have a non-empty intersection
(otherwise that connector is useless for the proposed query).

%

\section{Conclusions}
\label{sec:conclusions} We have described a method to represent
and solve the unicast/multicast QoS routing problem with the
combination of graph/hypergraph and SCLP programming: \emph{i)}
the best path found in this way corresponds to the best unicast
route distributing (for example) multimedia content  from the
source to the only receiver; \emph{ii)} the same considerations
are also valid for the best tree found over an \emph{and-or}
graph, since it corresponds to the best multicast distribution
tree towards all the receivers. The best path/tree optimizes
objectives regarding QoS performance, e.g. minimizing the global
bandwidth consumption or reducing the delay, and can satisfy
constraints on these metric values at the same time. The structure
of a {\em c-semiring} defines the algebraic framework to model the
costs of the links, and the SCLP framework describes and solves
the SCSP problem in a declarative fashion. Since several distinct
criteria must be all optimized (the costs of the arcs may include
multiple QoS metric values), the best route problem belongs to the
multi-criteria problem class, i.e. it can result in a
partially-ordered problem. Moreover we have seen also how to deal
with modality-based problems, relating them to preferences
connected to policy routing rules. Therefore, the model proposed
in this paper can be used to reason upon (and solve!) CBR, that
is, in general, a NP-Complete problem. 

In the future, we plan to enrich this framework by using
\emph{Soft Concurrent Constraint Programming} (SCCP)~\cite{sccp}
to handle the interactions among the routing devices and the
receivers, and, consequently, we would like to introduce new
``soft'' operations (e.g. a~\emph{retract} of a constraint) to
enable the release of the resources reserved by the receivers of
the communication, introducing a non-monotonic evolution of the
constraint store (which is not allowed in classical SCCP). A
second step should consist in further expanding the SCCP framework
with some simple primitives which allows to specify timing
constraints~\cite{dp}. Time critical aspects are essential to the
management of QoS, and, in general, when modelling the possible
interactions among distributed or concurrent systems. These
entities must continuously react to the inputs coming from the
environment and act in an appropriate manner.

A further extension to our QoS framework could be the introduction
of probabilistic metrics as the weight of the graph-links: we
could consider this value as the probability of packet loss on
that connection, or the probability of a connection existence
between two nodes in the network. In this case, the global
probability of existence of a path between two nodes $p$ and $v$
depends on the probability of all the possible different paths
connecting $p$ and $v$ in the graph. The problem is represented by
the composition of the different probabilities of these paths,
which cannot be easily modelled with a c-semiring, but we could
use the formulation given for \emph{semiring valuation} and
introduced in Sec.~\ref{sec:softCLP}, where the $+$ operator of
the c-semiring is non-idempotent.

At last, we will study if it is possible to represent the operators
used in MST and ST algorithms (see Sec.~\ref{sec:npcomplete})
with semiring structures.


\bibliographystyle{acmtrans}
\bibliography{biblio}

\end{document}